# New Candidates for Organic-rich Regions on Ceres

J. L. Rizos[1,2], J. M. Sunshine[2,3], R. T. Daly[4], A. Nathues[5], C. De Sanctis[6], A. Raponi[6], J. H. Pasckert[7], T. L. Farnham[2], J. Kloos[2], J. L. Ortiz[1]

1 - Instituto de Astrofísica de Andalucía, Spain (jlrizos@iac.es)

2 - University of Maryland - Department of Astronomy, USA

3 - University of Maryland - Department of Geology, USA

4 - Johns Hopkins University Applied Physics Laboratory, MD - USA

5 - MPI for Solar System Research, Germany

6 - Instituto Nazionale di Astrofisica, Italy

7 - Institut für Planetologie, Universität Münster, Germany

(Accepted for publication / PSJ - Oct 12, 2024)

## Abstract

We explore the spatial distribution of organics on Ceres using the visible and near-infrared data collected by the Dawn mission. We employ a spectral mixture analysis (SMA) approach to map organic materials within the Ernutet crater at the highest available spatial resolution, thereby revealing a discontinuous, granular distribution and a possible association with an ancient crater on which Ernutet has been superimposed. The SMA technique also helps us identify 11 new areas as potential sites for organics. These regions are predominantly located within craters or along their walls, resembling the distribution pattern observed in Ernutet, which implies a possible geological link with materials exposed from beneath the surface. In one of these candidate regions situated in the Yalode quadrangle, we detected the characteristic 3.4-micron absorption band in the infrared spectrum, indicative of organics and carbonates. By combining the spatial resolution of the Framing Camera data with the spectral resolution of the Visual and Infrared Imaging Spectrometer using SMA, we investigated the distribution of the 3.4-micron band in this quadrangle. The absorption pattern correlates with the Yalode/Urvara smooth material unit, which formed after significant impacts on Ceres. The association of organic-rich materials with complex and multiple large-impact events supports an endogenous origin for the organics on Ceres.

# 1. Introduction

Ceres, with a diameter of approximately 940 km, is the largest object in the main asteroid belt and located at a mean distance of 2.75 au from the Sun. Its unique physical and chemical properties set it apart from typical main belt bodies. In addition to its size, this dwarf planet is notable for being, after Earth, the most water-rich body in the inner solar system (Li & Castillo-Rogez, 2022). This composition has been inferred from its low density (2.16 g/cm$^3$), which suggests the presence of substantial amounts of internal water ice (McCord et al., 2005; Thomas et al., 2015; Dickson et al., 2022). Furthermore, based on the albedo and visible/IR spectral properties (a C-type), Ceres has been associated with carbonaceous chondrites (McSween et al., 2017), which are considered among the most primitive components of the solar system.

Given Ceres' scientific significance, NASA launched the Dawn spacecraft in September 2007 to explore both Ceres and the asteroid Vesta. Equipped with a Visible and Infrared Spectrometer (VIR) (De Sanctis et al., 2011), two Framing Cameras (FC1 and FC2) (Sierks et al., 2011), and a Gamma Ray and Neutron Detector (GRaND) (Prettyman et al., 2011), Dawn reached Ceres in 2015. The data acquired revealed a surface composed mainly of opaque materials, phyllosilicates, ammoniated bearing minerals, carbonates, water ice, and salts (e.g., McCord et al., 2018; De Sanctis et al., 2019).

In 2017**,** De Sanctis et al. identified aliphatic[1] organics near the Ernutet crater. This material was identified based on an absorption band at ~3.4 µm corresponding to CH group vibrations in the infrared spectra captured by VIR (see Fig. 1). Additionally, the organic-rich region exhibited a reddening of the spectral slope in the visible range, as observed by FC2, which distinguished it from the surrounding terrain (Nathues et al., 2016; Pieters et al., 2018). In addition to the Ernutet crater, three other regions have been suggested as potential locations for organics: the Inamahari crater (De Sanctis et al., 2017), the Occator crater (Raponi et al., 2021), and the most compelling candidate for organic material, the Urvara basin` (Nathues et al., 2022).

[1] An aliphatic organic is a compound containing carbon and hydrogen joined together in straight chains, branched chains, or non-aromatic rings. Examples of aliphatic are methane, propane, or asphaltite. The most significant characteristic of aliphatic compounds is that most of them are flammable.

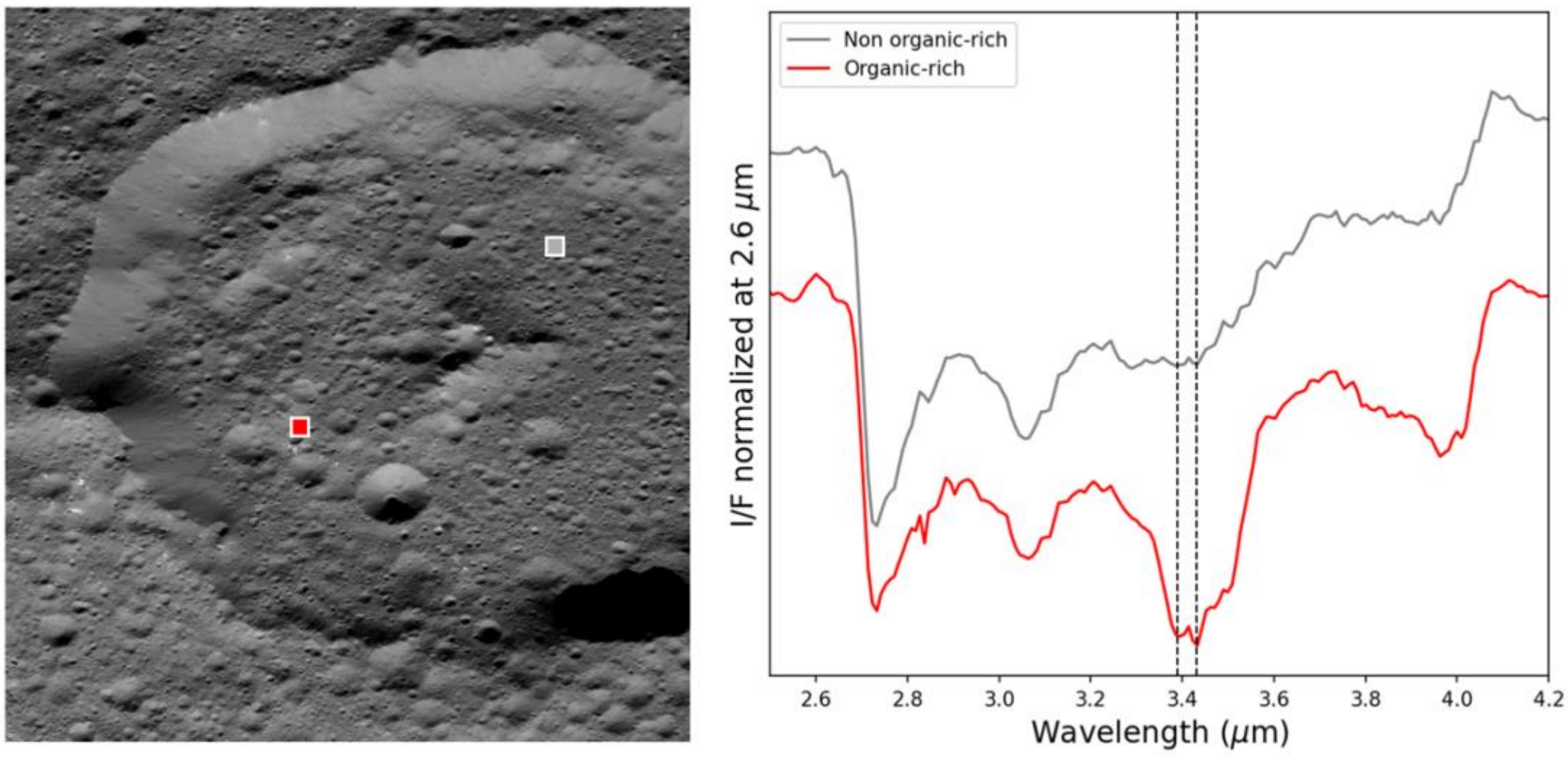

**Figure 1. The left panel displays an FC2 image showcasing the Ernutet crater, where organics have been detected. The red and gray squares correspond to organic-rich and non-organic-rich areas, respectively. On the right panel, the VIR spectra of these areas are presented, utilizing the same color scheme and with an offset for clarity. Within the organic-rich area, the red spectrum exhibits a distinctive absorption band near 3.4 µm. Ranging from 3.3 to 3.6 microns, the wavelengths display an asymmetrical shape with numerous sub-features, highlighted here by prominent black dashed vertical bars around 3.38 to 3.39 microns and 3.40 to 3.42 microns, indicative of aliphatic hydrocarbons (De Sanctis et al., 2017).**

Given its abundance of water ice and presence of organics, Ceres represents a possible example of objects that accreted into the terrestrial planets, delivering water and organics to the inner regions of the early solar system (De Sanctis et al., 2021). Moreover, the Dawn data provided evidence of the presence of a subsurface brine layer beneath the surface at the mantle-crust transition, which could be responsible for the ongoing geological activity observed on Ceres (e.g. Ruesch et al., 2019; Nathues et al., 2020; Raymond et al., 2020). Notably, the existence of brines and salty water places Ceres among the candidate ocean worlds with astrobiological implications.

An outstanding question revolves around the origin of Ceres' organic matter. On one hand, there are reasons to suspect that it could be exogenous material delivered via small organic-rich asteroids or comets (De Sanctis et al., 2017, Pieters et al., 2018, Nathues et al., 2023). This hypothesis requires the existence of an external source rich enough in organics, either from the main asteroid belt or the outer solar system to deposit these materials on Ceres. Furthermore, the physical conditions must be such that these organic materials have survived after impact, and the hypothesized source must be right enough in organics to explain the abundances seen on Ceres after accounting for dilution with native Ceres material in the ejecta (Bowling et al., 2020). In the solar system, organics have been identified in asteroids (Campins et al., 2010; Licandro et al., 2011), comets (Fran et al., 2016,) Jupiter trojans (Wong et al., 2024), moons of the giant planets (Tosi et al., 2024), and even in transneptunian objects (Emery et al., 2024a). If Ceres' organics were delivered by impacts, they would occur at large scales and be distributed randomly and globally. Moreover, Ceres lacks a substantial atmosphere and global magnetic field. Thus, the surface of Ceres is exposed to the full range of incident charged particle radiation, from low-energy solar wind ions to ultra-high-energy galactic cosmic rays (Nordheim et al., 2022). Therefore, those impacts must have been relatively recent. For instance, Marchi et al., (2019) estimated a C–H decay timescale between 10 and 100 Myr due to the vulnerability of the C–H bond to damage by ultraviolet and energetic proton radiation, although this timescale is dependent on the starting composition of H/C.

Another possibility that has been suggested is that the organics are endogenous materials that formed inside Ceres (De Sanctis et al., 2017; 2019; 2024). Under this model, there should be an internal reservoir of organics that, through some mechanism, is excavated. However, if we consider the hypothesis of an endogenous origin, it raises another question: why are the organics not widespread? Perhaps the answer is again related to the fact that aliphatic organics are vulnerable to solar interaction, with a progressive suppression of the 3.4 µm absorption band (Mennella et al., 2003; Godard et al., 2011, De Sanctis et al. 2024). If there is an internal reservoir, when covered, the organics would be protected from solar radiation, and only upon exposure would the degradation process begin. Thus, only those recently exposed organics would be identified, and the oldest exposed organics (less fresh) would have lost their characteristic absorption at 3.4 µm.

Bowling et al. (2020) have suggested an endogenous origin, asserting that their numerical models illustrate the inefficiency of an exogenic delivery of aliphatic organics. According to their models, the organic species would undergo thermal degradation and dilution upon mixing with target material in the ejecta blanket of a given crater. Nevertheless, other studies, although from a different approach, such as that presented by Daly et al. (2015) of hypervelocity impact experiments, demonstrate that within a specific range of impactor speed, size, and angle, and under certain porosity conditions, lightly shocked impactor material can survive and be deposited during impacts at typical Ceres impact speeds.

Additional follow-on studies have been conducted to help understand the spatial and compositional distribution of Ceres' organics. To date, experimental laboratory measurements have been conducted, comparing standard organics with VIR spectra, mapping potential organic-rich areas on Ceres, or comparing the concentration and distribution of hydrogen to determine the types of hydrogen-bearing species and how they were emplaced (Vinogradoff et al., 2021, Thangjam et al., 2018, Prettyman et al., 2019). However, none of these efforts have led to a definitive conclusion about the exact nature of these organic-rich materials. Moreover, while the geological context of organics found on Ernutet is relatively well defined, the origin of these organic-rich materials remains somewhat ambiguous, and it continues to be an open question.

Another challenge arises from the presence of carbonates distributed across the surface of Ceres, which complicates the identification of organic matter. Fig. 2 displays spectra of various carbonates thought to be present on Ceres' surface as documented by Carrozzo et al. in 2018. Specifically, it is presented the RELAB[2] spectra of calcite (specimen ID: CA-EAC-010), dolomite, (CB-EAC-003-A), rhodochrosite (CB-EAC-068-A), natrite (CB-EAC-079-A), antigorite (AT-TXH-002), and siderite (JB-JLB-E62-A). Certain carbonates, like dolomite or calcite, exhibit a pronounced absorption feature between 3.3 and 3.6 µm, which makes it challenging to identify organics using VIR spectrometer data unless organics are highly concentrated. Furthermore, while these organics are known to be aliphatic, given that their exact composition remains unknown despite laboratory attempts to replicate their features (Kaplan et al., 2018, De Sanctis et al., 2019, Vinogradoff et al., 2021), it is unable to rely on any other spectral feature outside the 3.4 µm in the infrared range that would unequivocally confirm their presence. In the visible range, Rousseau et al. (2020) conducted an in-depth study of Ceres' surface, concluding that there is no correlation between the spectral slope in the visible and the presence of carbonates.

[2] The PDS Geosciences Node Spectral Library is a collection of measurements of Earth, lunar, and meteorite materials to be used to compare to flight measurements. https://pds-speclib.rsl.wustl.edu/

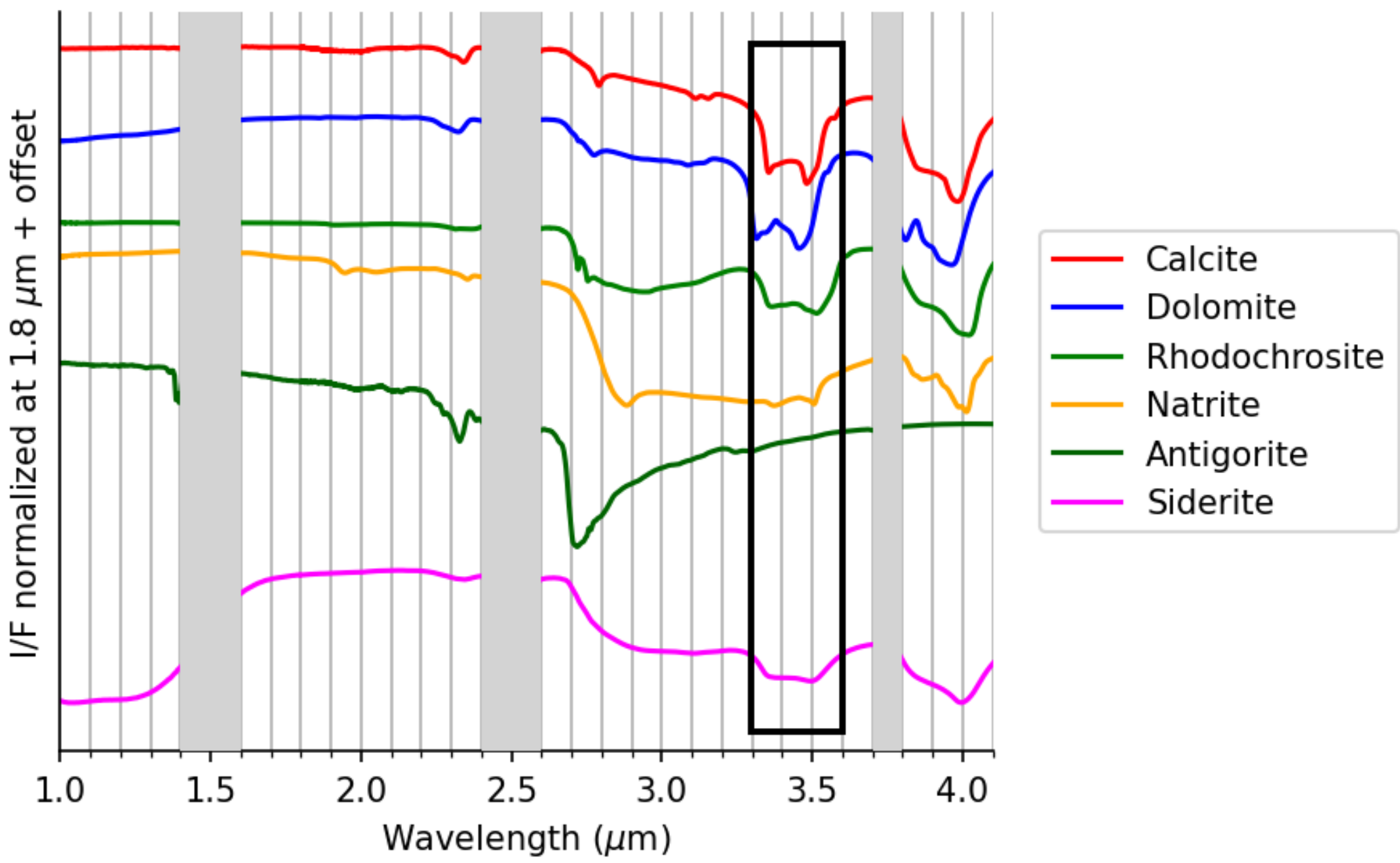

**Figure 2. Infrared spectra obtained from the RELAB database for carbonates that could be present on the surface of Ceres. Most carbonates include an absorption band between 3.3 and 3.6 µm, complicating the identification of the 3.4 band characteristic of aliphatic organics when both are present. Vertical grey bars represent regions of the detector affected by the filter junctions (De Sanctis et al., 2011).**

Additionally, it has been suggested that phyllosilicates (2.7 µm absorption band) could provide protection for organics against space weathering effects. Poch et al. (2015) demonstrated a significant photoprotective effect of clay minerals, which efficiently preserve some organic molecules against UV irradiation, making their presence and mixing conditions potentially relevant. Therefore, to ensure an accurate interpretation, it is essential to conduct a multi-focus study where the spectral absorptions are evaluated, their relationships with other spectral features observed on the surface, as well as their morphology and distribution.

Confirmation of additional regions containing organic material without traces or vestiges of impacts on the distribution would lend support to the endogenous hypothesis. On the contrary, the identification of a morphology compatible with an impact in the Ernutet crater, and a total absence of organics on the rest of the surface, would lead to considering the exogenous hypothesis more plausible. To further investigate these possibilities, we employ a new approach to explore the surface of Ceres in search of organic materials and to analyze its distribution at the highest possible spatial resolution. Dawn's Framing Camera color filters enable high spatial resolution imaging but with low spectral resolution. Conversely, the VIR imaging spectrometer provides high-resolution spectra that facilitate compositional characterization of the surface but at a lower spatial resolution than the camera. In this study, we combine the high spatial resolution provided by the FC2 camera with the high spectral resolution of the VIR spectrometer using an innovative approach developed for Dawn data of Vesta (Cheek et al., 2020) that allows us to analyze the distribution of organics on Ceres with unprecedented detail.

To begin with, we employ a spectral mixture analysis (SMA) algorithm to investigate the Ernutet crater region, thereby validating our method and facilitating a morphological analysis of this organic-rich area. Using SMA, we systematically investigate the entire surface of Ceres using FC2 images, aiming to identify candidate regions potentially rich in organic material. As a result, we identify 11 promising candidates exhibiting characteristics such as spectral reddening, which suggest the presence of organics. We then thoroughly examine the spectral properties of these candidate regions using the high-resolution spectra obtained by the VIR instrument. This analysis confirms the presence of the 3.4 µm absorption band in 1 out of 11 identified regions, located within the Yalode quadrangle. However, the presence of carbonates in this quadrangle complicates an unambiguous identification of organics. To provide a broader context for this finding, we compare the carbonates in this region with the rest of the main carbonated spots on Ceres, and in turn, we generate a map combining the spatial resolution of FC2 and the spectral resolution of VIR. This enables us to assess the relative abundance and distribution of the carbonates and possible organic materials within the region and conduct a comparative analysis with the geological units identified by Crown et al. (2018).

# 2. Data analysis

## 2.1 Data preparation

*FC2 camera*: This camera images in seven color filters centered at 438, 555, 653, 749, 829, 917, and 965 nm (Sierks et al., 2011). PDS[3] images (Nathues et al., 2015) are processed and converted into multispectral cubes to be further processed with ISIS[4]. A thorough description of radiometric calibration is provided by Kovács et al. (2024). SPICE kernels from the DAWN spacecraft (Krening et al., 2012) are then used to georeference the images. Following this step, we apply photometric corrections using the parameter-less Akimov model as the disk function, coupled with a second-order polynomial phase function, as detailed in Rizos et al. (2019). Photometric backplanes needed for this task are obtained using a digital terrain model derived from the FC2 HAMO (High Altitude Mapping Orbit) images, with a pixel scale of 136.7 meters per pixel, generated using stereo photogrammetry methods by Roatsch et al., (2016) and can be found in the PDS. The resulting product is projected using the equidistant cylindrical projection, and the seven color filters are aligned with each other using an iterative adaptative least squares correlation algorithm (Gruen et al., 1985). Finally, a global mosaic is assembled by combining the observed subregions.

*VIR spectrometer*: The VIR imaging spectrometer is composed of two channels: the VIS channel covers the spectral range 0.25–1 µm, with a spectral sampling of ~2 nm, and the IR channel covers the spectral range 1–5 µm, with a spectral sampling of ~10 nm (De Sanctis et al., 2011). Here we focus on data from the IR channels whose spectral range includes the diagnostic absorptions of organics and carbonates. PDS VIR data (De Sanctis et al., 2015) are affected by the odd-even effect and systematic artifacts. These have been corrected according to the method implemented by Carrozzo et al., (2016). Residual slope effects have been corrected as discussed in Raponi et al. (2023). The observed spectra of Ceres' surface are affected by thermal emissions longward of 3.2 µm, which hides the absorption bands and prevents comparison with laboratory reflectance spectra of analogous materials. We applied the thermal removal algorithm of Raponi et al. (2019) to generate reflectance data.

[3] https://pdssbn.astro.umd.edu/data_sb/missions/dawn/index.shtml

[4] ISIS is a powerful software tool specifically designed for manipulating imagery acquired by NASA planetary missions. It provides a comprehensive set of functionalities that enable the analysis of three-dimensional cubes generated by imaging spectrometers. With ISIS, users can project images and create detailed mosaics. More information about ISIS can be found at https://isis.astrogeology.usgs.gov.

## 2.2 Spectral Mixing Analysis

*Abundance maps:* In this analysis, we employ Spectral Mixing Analysis (SMA) as has been used to successfully analyze the Dawn FC and VIR data of Vesta (Cheek & Sunshine 2020). SMA, developed by Adams & Gillespie (2006), is a method in which a mixing model converts n-spectral images (where *n* is the number of wavelengths) into maps of a few representative scene endmembers. Therefore, endmembers may be spectra of pure materials or spectra of mixtures of materials. This is a technique that depends on the remote-sensing scale. For example, when observing the Earth's surface at a scale of tens of meters, two spectral endmembers could be defined: one for the soil and another for general vegetation. However, if we approach scales of the order of centimeters, we will need to define a different type of endmember in which each one represents a specific type of vegetation or soil.

The first step in SMA involves a preliminary identification of representative endmembers of a surface at the observed scale. Here we used a combination of band ratios, spectral clustering, and principal component analysis the help identify endmembers. These techniques allow us to distinguish regions with well-differentiated spectral behaviors. These preliminary endmembers are used for initial mixing models and refined based on the results. Once endmembers are defined, we calculate the proportion or fractional abundance, which ranges between 0 (when the endmember is not present) and 1 (when the endmember represents 100% of the mixture) of each endmember in each pixel. Thus, in each pixel, we consider that the measured reflectance at the *i*-th color filter ($R_i$) is a linear combination of the reflectances from each sub-pixel spectral endmember ($R_{Xi}$), plus an error term (Eq. 1).

$$\left.\begin{array}{c} R_1 = F_A * R_{A1} + F_B * R_{B1} + F_C * R_{C1} + error_1 \\ R_2 = F_A * R_{A2} + F_B * R_{B2} + F_C * R_{C2} + error_2 \\ \cdots \\ R_7 = F_A * R_{A7} + F_B * R_{B7} + F_C * R_{C7} + error_7 \end{array}\right\} \quad (1)$$

We constrain the result so that the sum of all the abundances ($F_X$) is one (100%). However, we do not constrain the abundance of each endmember between 0 and 1, which would be the most realistic from a physical point of view. Instead, we allow for quantities greater than 1 (*superpositives*), and quantities less than 0 (*negatives*). In allowing for such physically unrealistic results, areas that are poorly fit by the current endmember can be identified, which supports the identification of more appropriate or additional endmembers. Moreover, in SMA studies it is convenient to include a shade endmember that represents shadows and the effect of shading. It helps stabilize the SMA solution so that true compositional variations can be modeled with varying proportions of the endmembers, independent of any accompanying brightness variations arising from illumination or topography (Adams et al., 1995; Adams and Gillespie 2006; Cheek and Sunshine 2020).

As an additional assessment, for each pixel and wavelength, we calculate the error as a sum of the total root mean square (RMS) over all wavelengths, enabling us to generate an error image that considers the appropriateness of the chosen endmembers through a global inspection of our area. If a region with a high RMS error appears in the error image that is unrelated to known artifacts, e.g., resulting from the mosaic construction process, it suggests that the initial choice of endmembers needs refinement. Thus, while the method primarily relies on data analysis techniques, its final stage requires human supervision. It should be noted that linear mixing is assumed under spectral mixture analysis (SMA), which is not strictly true. Nonetheless, the resulting errors are negligible for our purposes (Cheek & Sunshine 2020).

<u>Extrapolated dataset</u>: Following the methods of Cheek and Sunshine (2020), the next step in our analysis involves combining the spatial resolution of FC2 with the spectral resolution of VIR to create an 'extrapolated dataset'. Essentially, this step entails determining the VIR spectral signatures of the endmembers previously identified with FC2 in VIR, now called 'hyperspectral endmembers.' Subsequently, for each FC2 pixel, we replace the original FC2 endmembers with these hyperspectral endmembers mixed with the same fractional abundances determined from the FC2 data. The assumption is that the physical endmembers and their abundances are the same in the FC2 and VIR.

To achieve this, first, we select a sufficient number of locations (n) within our scene that represent spectral diversity. We then model the spectra of all locations to determine the VIR hyperspectral endmembers given the abundance of each endmember based on the previously obtained FC2 SMA solution. Next, we subtract the shade from the VIR spectrum, obtaining the shadeless spectra. Finally, we conduct a least-squares fit for all locations, assuming that this shadeless spectrum is the sum of the hyperspectral endmembers plus an error (Eq. 2)

$$\left.\begin{array}{c} R_1(\lambda_i) = F_{A-1} * R'_A(\lambda_i) + F_{B-1} * R'_B(\lambda_i) + error_1(\lambda_i) \\ R_2(\lambda_i) = F_{A-2} * R'_A(\lambda_i) + F_{B-2} * R'_B(\lambda_i) + error_2(\lambda_i) \\ \ldots \\ R_n(\lambda_i) = F_{A-n} * R'_A(\lambda_i) + F_{B-n} * R'_B(\lambda_i) + error_n\,(\lambda_i) \end{array}\right\} \quad (2)$$

in which $R_X(\lambda_i)$ represent the *shadeless* VIR reflectance at location *X* for the *ith*-wavelength, $F_{Y-X}$ is the abundance of the *Y*-endmember at the *X* location, and $R'_Y(\lambda_i)$ – the unknown to be solved for– represents the reflectance of the hyperspectral endmembers at the *ith*-wavelength.

The extrapolated dataset is now built by summing for each FC2 pixel the fractional abundances multiplied by their hyperspectral endmember VIR-derived reflectance over the whole FC2 scene.

Under SMA, it is assumed that the scene can be described by only a few of endmembers, and the spectral behavior identified in the visible range can be extrapolated to the infrared (Cheek and Sunshine, 2020). This latter assumption is reasonable given that Pieters et al. (2018) found VIR is spatially correlated with a distinctive, red-sloped continuum measured by FC2 in the visible. Thus, while SMA and the extrapolated datasets are excellent tools for jointly exploring the best available high-spectral and high-spatial data for Ceres, compositional interpretations must be carefully evaluated using the measured VIR data.

# 3. Results

## 3.1 Endmembers identification

Our first focus is identifying organics on the surface of Ceres as observed from the FC2. To test our approach, we initially selected a broad region centered on the Ernutet crater, encompassing a sufficiently large area that includes not only the previously identified organics-rich zones (De Sanctis et al., 2017) but also the surrounding area. This reference mosaic is constructed using FC2 color images obtained during the High-Altitude Mapping Orbit (HAMO) phase of the Dawn mission, with a pixel scale of approximately ~136 meters per pixel and a phase angle of ~50°.

In this analysis, we identified three endmembers in this region, which are presented together in Fig. 3 as an RGB map. This map combines an endmember for the region where there are organics (red), a background endmember representing the average behavior of Ceres (green), and the shade endmember (blue). Here the shade is an endmember spectrum having 0.01 values in all channels that represent shadows and the effect of shading.

Note that the organic endmember does not describe a pure organic area but rather the most characteristic and well-differentiated region with organics (plus other compounds) as seen at the spatial scale of FC2. Therefore, if the fraction abundance of the organic endmember in a given pixel is 1 (100%), it does not mean that there are 100% organics, but there is 100% of that endmember. To avoid possible confusion between the pure compound and the endmember, we henceforth use the letters O, B, and S to denote the organics, background, and shade endmembers, respectively. The fraction abundance maps for each endmember are displayed in Appendix Fig. A.

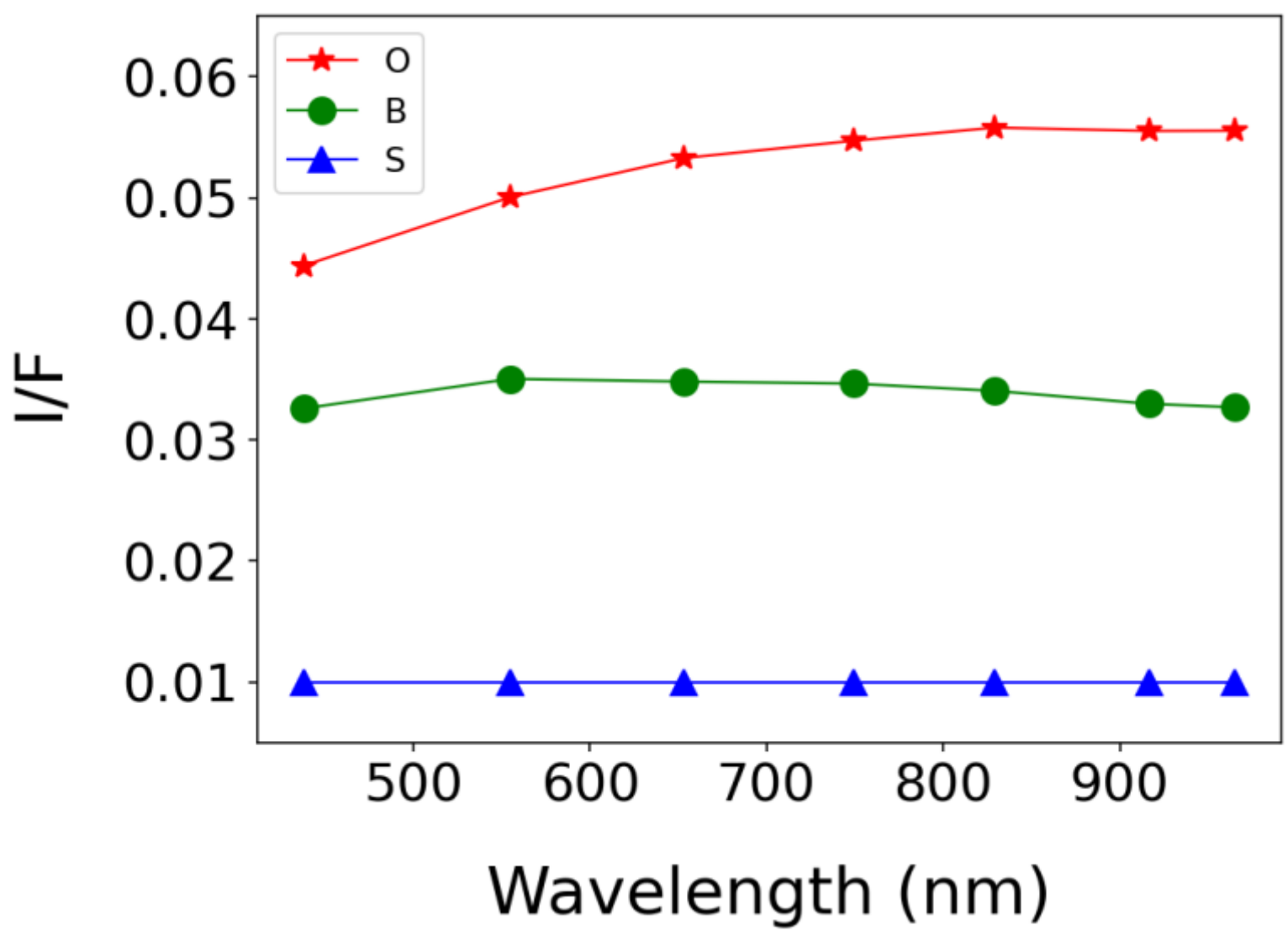

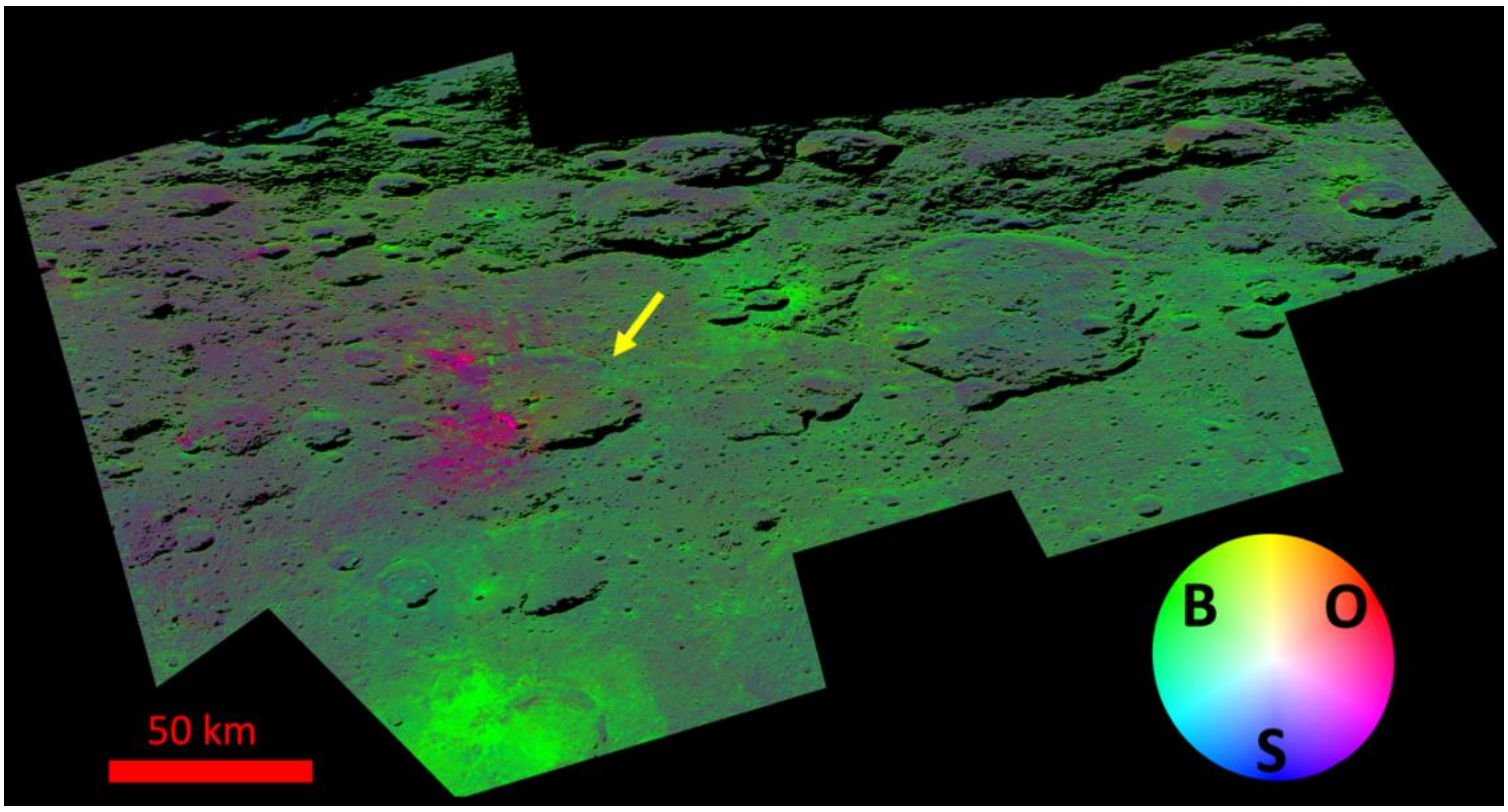

**Figure 3. Top panel: FC2 spectral endmember O (organic-rich), B (global Ceres background), and S (shade), respectively. Bottom panel: RGB map where the abundance of the endmember O is represented in red, the endmember B in green, and the S in blue. The pixel scale of this mosaic is ~136 m/pixel and the images were acquired with a phase value of ~50. Ernutet is the central crater (pointed by the yellow arrow) of the scene in which the organics are visible.**

## 3.2 Ernutet region

Once the endmembers of the global scene have been identified, we focus on the subregion with the highest concentration of aliphatic organics in Ernutet, applying SMA. For this purpose, we used FC2 images from phase XMO6, which are the color filter images with the highest spatial scale on Ernutet (36 m/pixel), and the result is shown in Fig. 4. The fraction abundance maps for all endmembers are displayed in Appendix Fig. B.

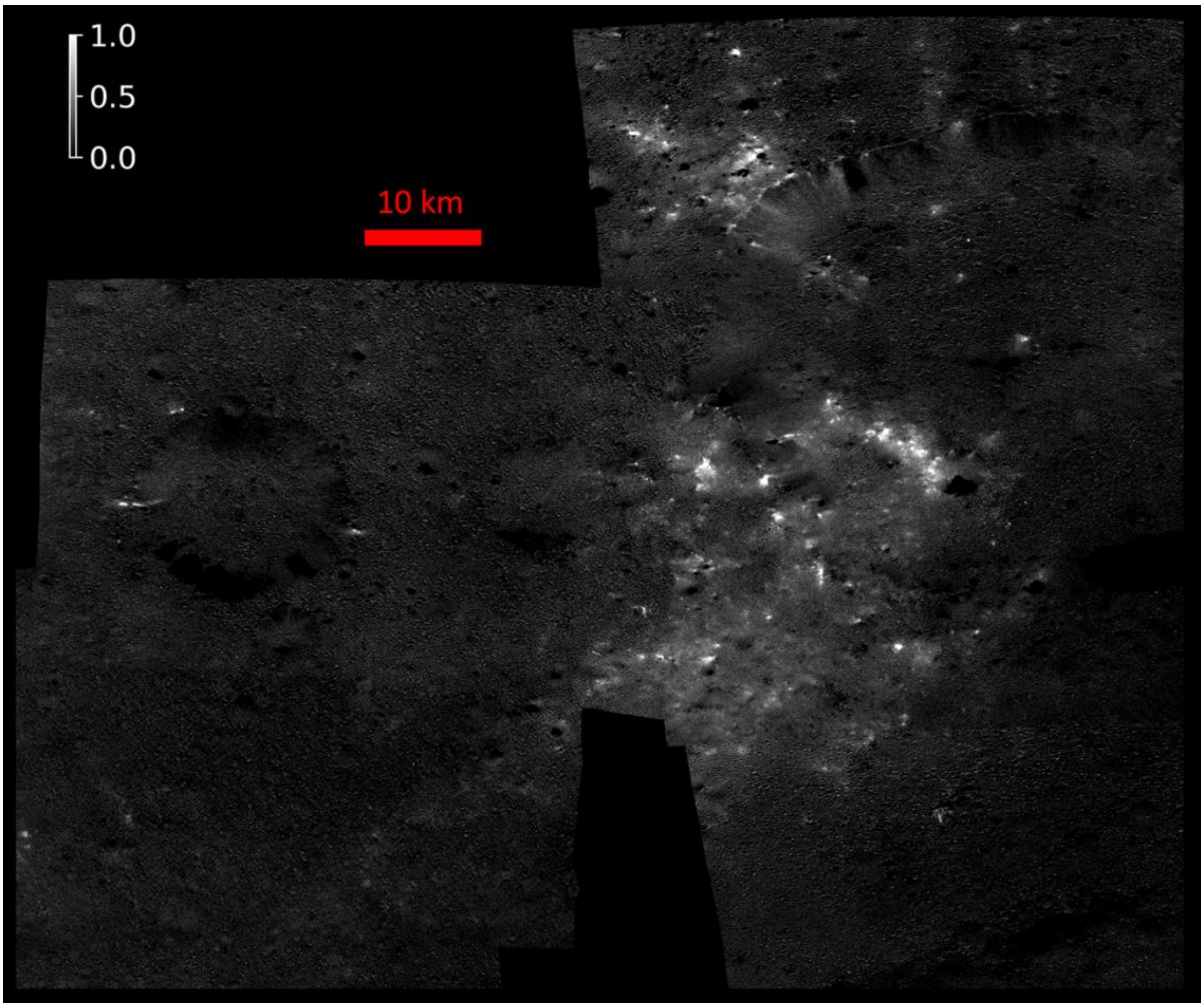

**Figure 4. Abundance map of the endmember O (representative of aliphatic organics) in the Ernutet crater with a pixel scale of 36 m/pixel (XMO6 phase images).**

The result is consistent with the analyses and the distribution found by other authors (Pieters et al., 2018; De Sanctis et al., 2019), thus validating our new approach. However, unlike those other works, here we combine the highest spatial scale available from FC2 with the depth of the absorption band at 3.4 microns provided by VIR. We observe that the organic-rich areas do not exhibit the expected morphological characteristics following an impact. Typically, an impact would result in material accumulations within the crater, a well-defined distribution of ejecta (debris ejected from the crater), and the formation of secondary craters or other secondary surface modifications. These secondary features often include rays or chains of craters created by re-impacting debris. In contrast, our observations reveal that the organic material is distributed indiscriminately. It appears both within and outside the crater, including on the walls. Qualitatively, we note that rather than seeing concentrated deposits, the organic material is spread out in a granular distribution with highly localized areas showing high abundances. These areas stand out against a more diffuse background.

To better understand the geologic context of the distribution of these bright spots, in Fig. 5a, we placed a series of color-coded stars on a map overlaying the organic compounds (red layer) onto an FC2 mosaic (550 nm). The distribution of the brightest spots –and therefore, the higher concentration of aliphatic organics – suggests a circular pattern starting from the rim of Ernutet (blue star) to the farthest region (red star). Overlaying these spots on another image taken with the clear filter from a different viewing geometry (Fig. 5b), reveals that these organics are located within an adjacent and what is possibly an ancient crater –indicated by the yellow dashed line– than Ernutet. This crater is interpreted as an older structure based on its rim morphology and the principle of superposition. First, the crater displays a shallow rim, a feature commonly associated with older structures, as crater rims tend to erode and lose prominence over time. Within this structure, we observe smaller, well-defined craters, with Ernutet clearly superimposed on top. For a better visualization of this structure, refer to Fig. 3 of the digital terrain model in Pasckert et al. (2018).

In fact, we observe that this southwest diffuse organic spot is located within this older structure. De Sanctis et al. (2019), based on Ceres' internal evolution, have proposed several upwelling mechanisms to justify the hypothesis of an endogenous nature of these organics. This distribution could indicate that the ancient impact played a significant role in enhancing these upwelling mechanisms. It is conceivable that Ernutet may have exposed deeper material here, which was initially brought closer to the surface by the older, more ancient crater. Furthermore, while Ernutet's ejecta completely overlaid this older crater, subsequent smaller impacts could also uncovered the material beneath.

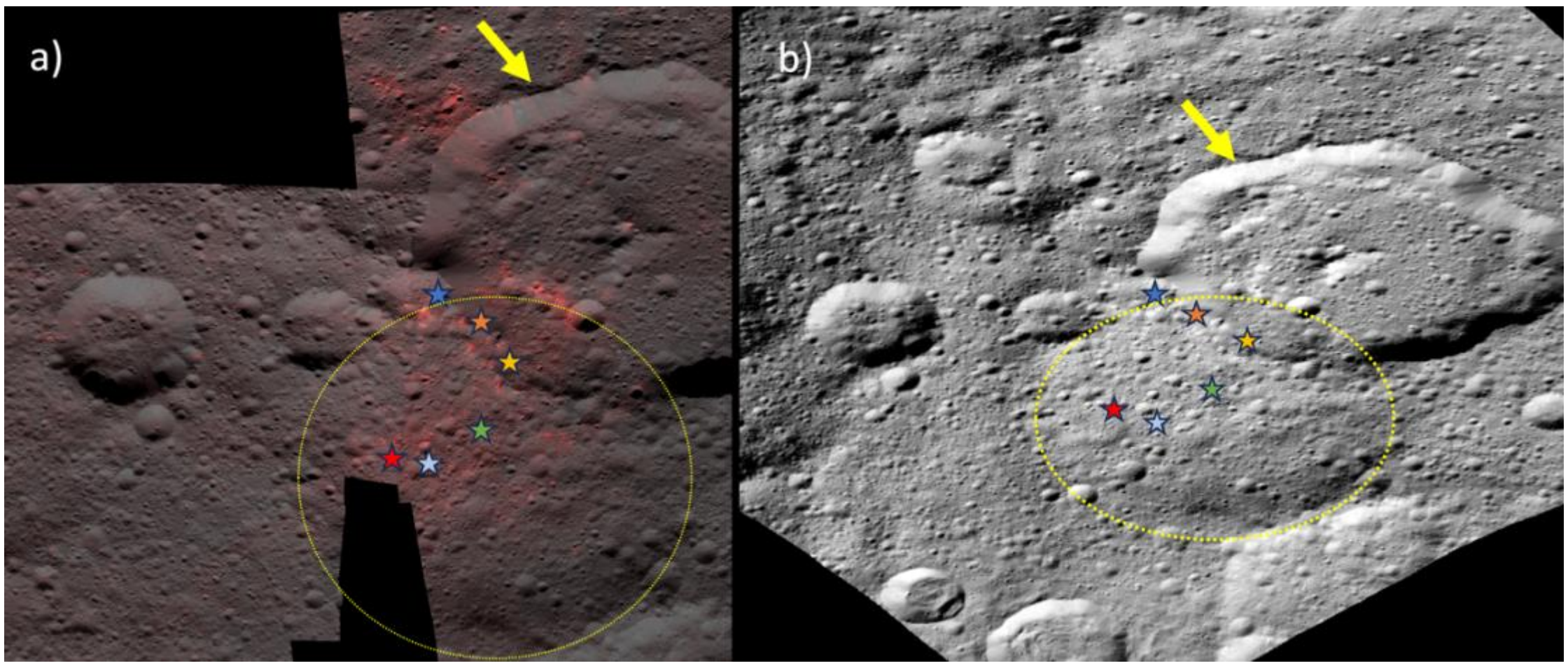

**Figure 5. a) Areas rich in aliphatic organics (red) obtained after applying SMA (Fig 2), overlaid on a F2 mosaic with XMO images. The Ernutet crater is pointed out by the yellow arrow. The color-coded stars indicate areas of higher concentration of the organics-rich endmember in the outer region in the southwest of Ernutet. The yellow line marks the rim of the ancient crater on which Ernutet is superposed. b) Same as in (a) but using a clear filter image with a different viewing geometry, thereby allowing for a better appreciation of the spatial distribution relative to the older crater. The higher concentrations of the organic-rich endmember within the degraded crater could be related to the unearthing of deeper-seated material by the ancient impact. Additionally, subsequent smaller impacts may have further exposed the underlying material as Ernutet's ejecta concealed the older crater entirely.**

## 3.3 Global survey

Using the endmembers identified near Ernutet, we run a pipeline to systematically search the entire surface of Ceres for additional regions that are potentially organic-rich. We follow the same procedure used for the mosaic in Fig. 2, using Cycle 1 and 2 of HAMO FC2 color images. We chose this set of images because they cover the entire surface, and they present the same pixel scale as our reference mosaic (~140 m/pixel), with a similar phase value (~50°). Therefore, we avoid spatial scale differences and the spectral reddening effect with phase angle (Kitazato et al., 2008; Reddy et al., 2012), which could complicate the identification.

The surface is divided into a total of 37 sub-regions, each of which is individually examined. Our criterion for considering a region as a potential candidate for organic matter is the presence of a spatially coherent group —in endmember O— with a fraction abundance exceeding 80%. Finally, we examine the VIR spectra to corroborate or dismiss the presence of organics by analyzing the presence or absence of the characteristic 3.4 µm absorption band.

The map in Fig. 5 shows the regions where we have identified potential organic sites using our SMA methodology. In total, we have found 13 spots. Regions previously identified by other researchers and re-confirmed using our technique —validating our methodology— are highlighted in yellow. This includes the Ernutet region as identified by De Sanctis et al. (2017) and the Urvara basin as identified by Nathues et al. (2022). The green stars indicate candidate regions for organic matter, which are not confirmed because they lack unambiguous absorptions at 3.4 µm in the VIR data. The most interesting case, indicated in Fig. 6 with a red star, is the candidate 12, where we found a spectral feature at 3.4 µm. This region is in the Yalode quadrangle (Ac-14), situated between 21° to 66°S and 270° to 360°E.

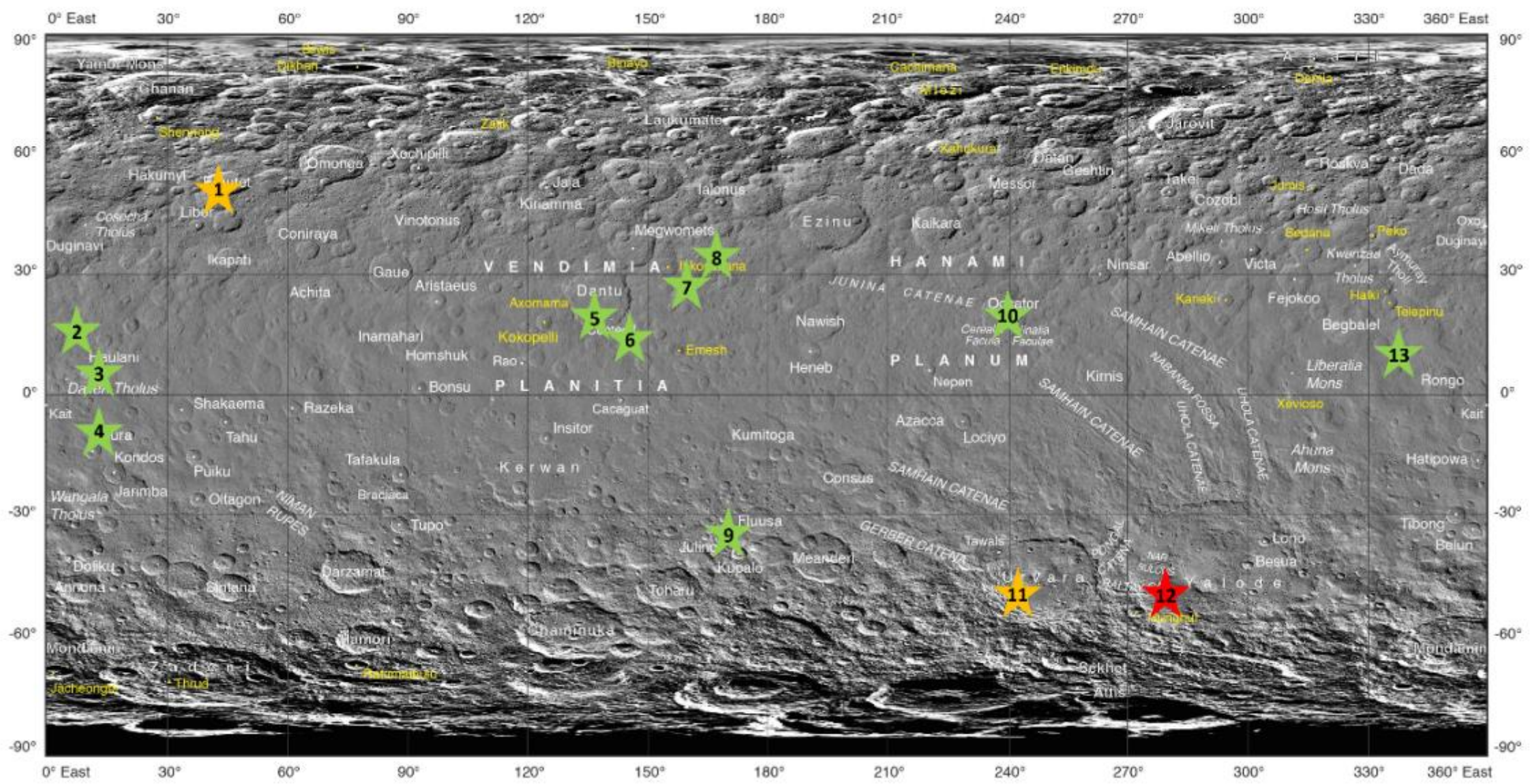

**Figure 6. Ceres surface global map in which we indicate the candidate organic regions. Yellow stars indicate the Ernutet crater (1) and the Urvara basin (11) where organics have previously been identified (De Sanctis et al., 2011, Nathues et al., 2022). In red is the region where we have identified a new organic region, located in the Yalode quadrangle. Green stars mark locations where SMA suggested that organics may exist, but can not be confirmed because they lack 3.4 µm absorptions. Credits for background map: NASA/JPL-Caltech/UCLA/MPS/DLR/IDA**

Fig. 7 displays the RGB images of 10 new candidates (the Yalode quadrangle candidate is covered in the next section), adopting the same color criterion as in Ernutet: red for endmember O, green for endmember B, and blue for endmember S. Most of these candidates are situated within craters or along their walls, mirroring the pattern observed in Ernutet, suggesting a potential geologic association with materials that have been exposed from below. There is a noticeable concentration of these non-confirmed candidates in the equatorial region, primarily within the latitude range of (-40°, 40°). Furthermore, we can identify two spatial clusters among these candidates. The first group includes candidates 2, 3, 4, and 13, ranging from 330° to 90° longitude. The second group consists of candidates 5, 6, 7, and 8, spanning from 120° to 180° longitude.

Candidate 10 (Fig. 7i) corresponds to the interior and walls of the Occator crater. This region has been proposed by Raponi et al. (2021) as a candidate for the presence of organic material after analyzing the spectra in the visible channel of VIR. They determined that the red spectrum detected was similar to those observed in the organic-rich spectra coming from the Ernutet region.

Given that several analyses of aliphatic organics (Mennella et al., 2003; Godard et al., 2011) predict a progressive suppression of the absorption band at 3.4 µm due to solar interaction, we cannot entirely rule out the presence of aliphatic organics in these regions based on the lack of this band. Furthermore, it is also possible that aliphatic organics are present and have characteristic absorption bands but at lower concentrations that may not be detectable at the spatial scale of the VIR observations.

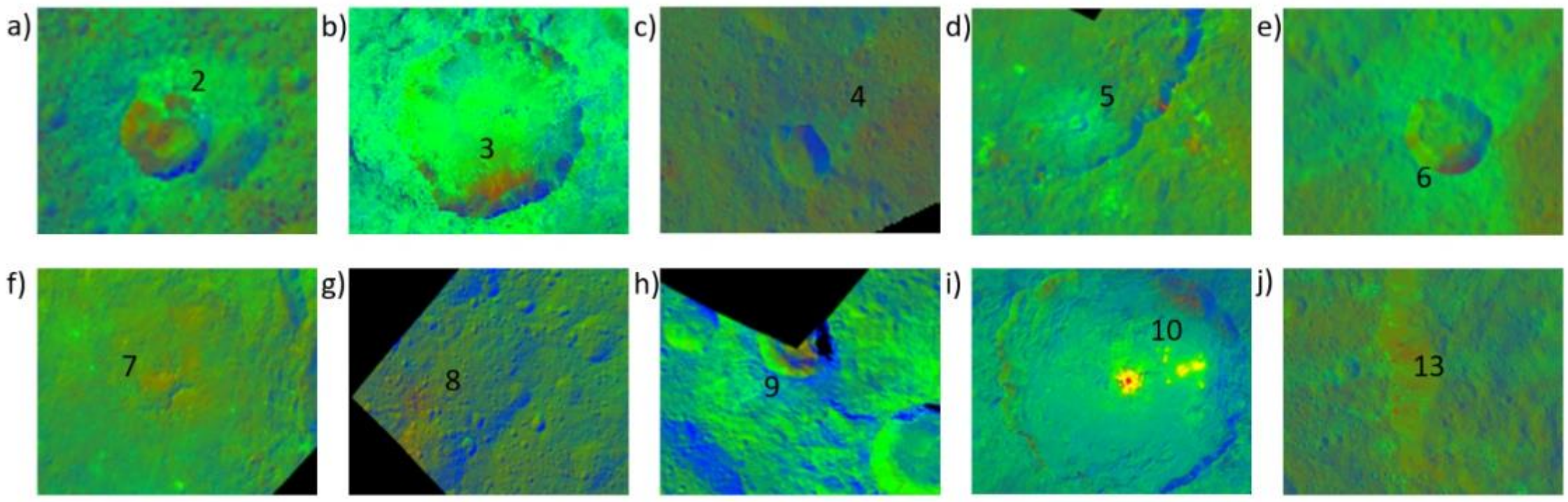

**Figure 7. Regions candidates to harbor organics according to the SMA algorithm and FC2 data. The images are oriented, north and east pointing up and to the left respectively. Except for candidates 4, and 8 (c & g), all are found either inside or on the wall of craters. However, the analyses carried out with the VIR spectrometer are not conclusive and thus the presence of organics cannot be confirmed in these ten locations. The colors are as in Fig. 3.**

It is worth noting that it is not possible to attribute to our candidates a purely non-organic carbonate or silicate composition, which are the predominant materials on the surface of Ceres. As an example, our candidate number 3 (Fig. 7b) is the well-studied Haulani crater, which was analyzed among other works by Tosi et al. (2018). They confirm that the largest concentration of carbonates is in the western inner wall and the northwestern rim of the crater, while here we find the candidate organic area in the south of the crater (i.e., not where carbonates were reported by Tosi et al., 2018). Moreover, Rousseau et al. (2020) confirmed that no obvious correlation is observed between visible spectral slopes and the distribution of carbonates around Haulani.

On the other hand, unlike carbonates, there is a correlation between Mg-phyllosilicates and the 480−800 nm slope, and between NH4-phyllosilicates and the 405−465 nm slope on Ceres' surface (Rousseau et al., 2020). However, phyllosilicates do not appear to be responsible either for the candidates identified with our algorithm. Using the Haulani crater again as an illustration, the phyllosilicate abundance in the interior of Haulani has been studied by Ammannito et al. (2016), and there is no association with our identified candidate organic area.

## 3.4 The Yalode candidate

The SMA conducted in region 12, situated between the Urvara and Yalode basins, reveals several concentrated groups and an extensive region that are organic candidates. This region falls within the Yalode quadrangle, located in the southern hemisphere of Ceres.

In Fig. 8a a clear filter FC2 mosaic of this region is presented. The red lines delineate the area covered by the seven color filters employed in this analysis. It is contiguous with the area indicated by Nathues et al. (2022) as a potential candidate for organic material (marked by a red arrow). Fig. 8b shows the resulting RGB map of SMA endmembers where the most prominent bright spots are labeled (white circles) as BS1, BS2, and BS3, with extents of 474, 2204, and 578 meters, respectively. The fraction abundance maps for each endmember are displayed in Appendix Fig. C. This scene includes numerous craters surrounded by ejecta, with a diffuse layer of the endmember O observed in the intervening area. The leftmost crater in the mosaic, characterized by a significant concentration of the endmember O, corresponds to the rim of the Urvara basin. The presence of a catena, channels, and various geological features, indicates that the candidate organic-rich areas are in a region that has experienced significant geological activity in the past.

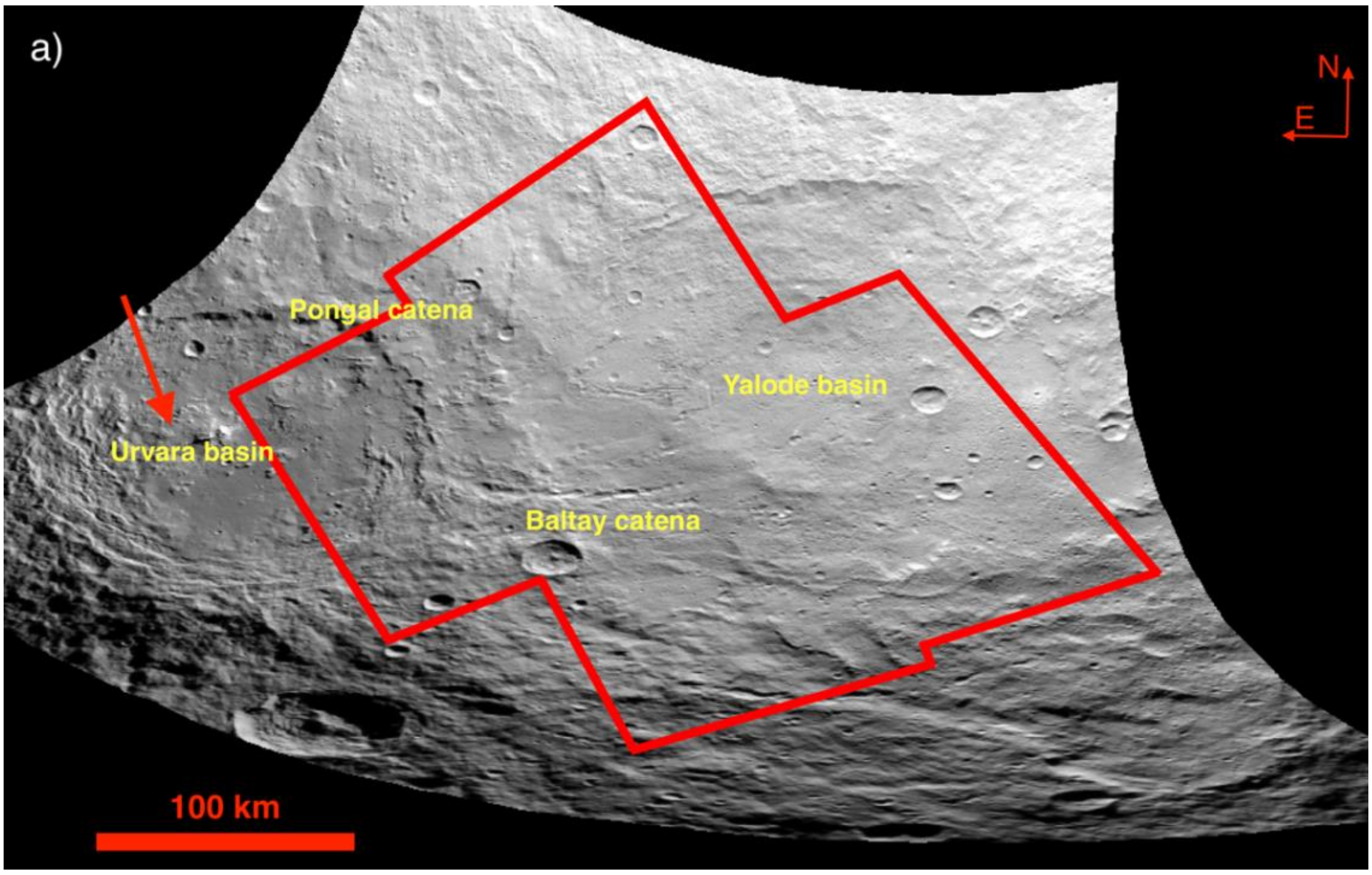

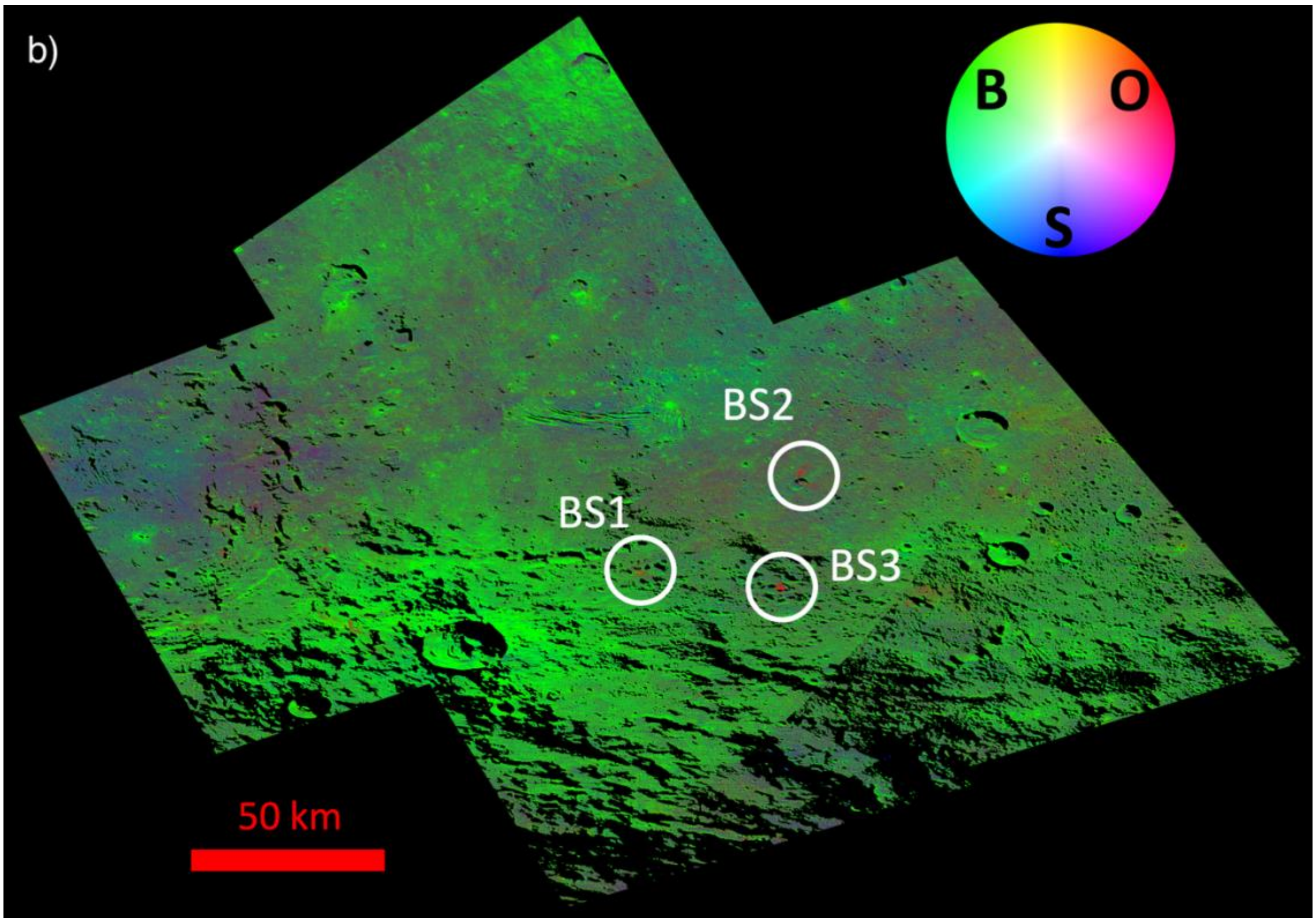

**Figure 8. a) Mosaic built from FC2 clear filter images of the region (centered at Yalode quadrangle) where organics candidates are suggested. The region outlined in red delimits the area of the FC2 color filters used in this work. B) RGB map of Yalode quadrangle where the O endmember has been identified (~136m/pixel). The abundance of the O endmember is represented in red, the B endmember in green, and the S in blue. Bright spots BS1, BS2, and BS3 show the highest abundance of potential organics, but there are also potential organics scattered in the layer over the adjacent north area.**

Fig. 9 compares the FC2 spectra obtained from the three prominent bright spots to other regions on Ceres. These spots exhibit a visible spectral shape similar to the reddish one observed in the Ernutet organic region. In the wavelength range of 438–653 nm, the spectra of these bright spots are virtually identical to those of aliphatic organics. In the range of 653–965 nm, the spectra of these bright spots appear slightly redder than the background, although not to the same extent as observed in the case of Ernutet. If the reason for this redness is the presence of aliphatic organics, it suggests that the organic concentration of the 3 spots is lower than in Ernutet.

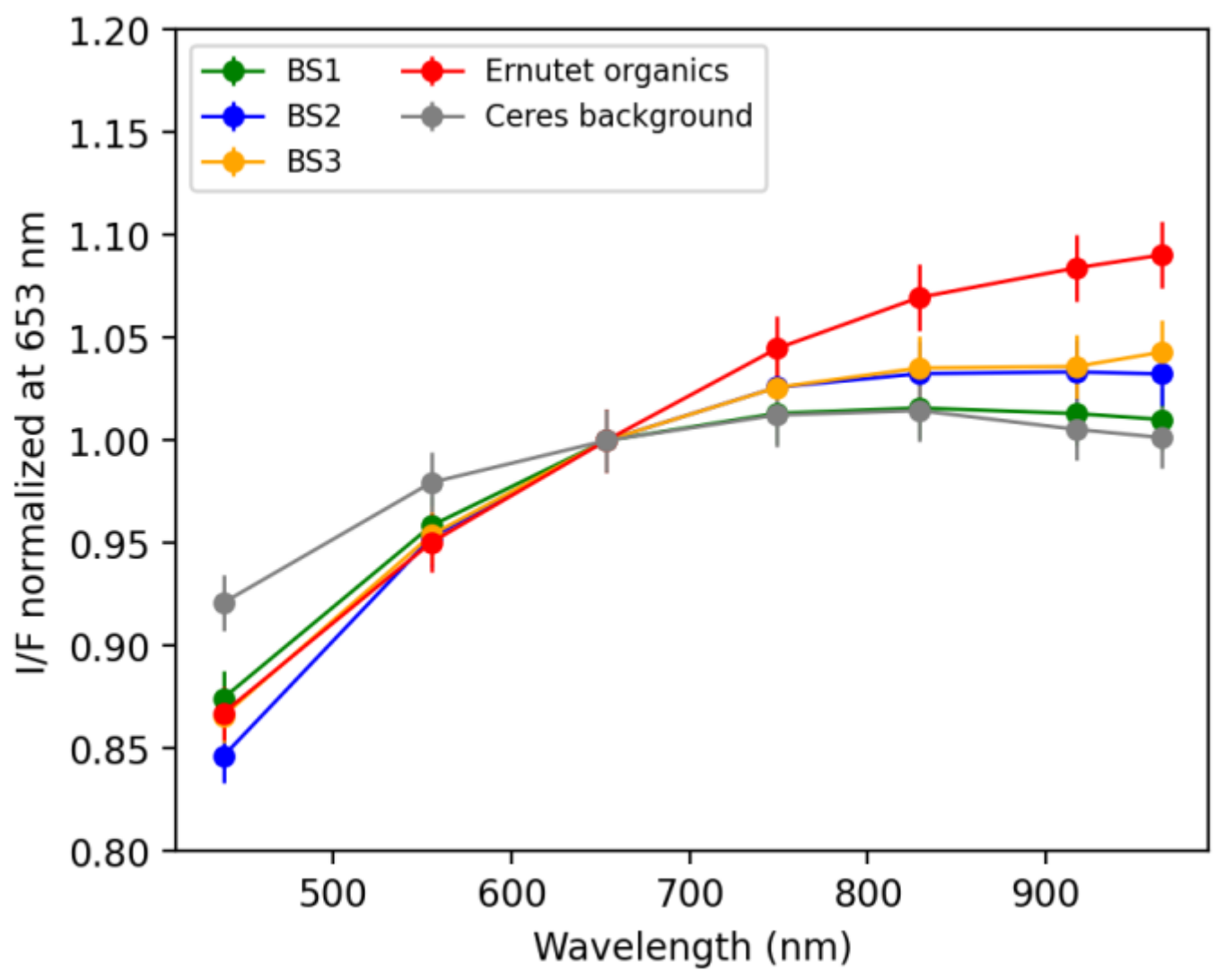

**Figure 9. FC2 spectra of the bright spots identified in Yalode, along with the average spectrum of Ceres (grey) and the Ernutet organics spectrum (red). All spectra are normalized at 653 nm.**

As seen in Fig. 10, BS1 and BS3 are amorphous features with an extension in its longest parts of 474 and 578 meters, respectively. While BS3 is between a cluster of small craters, BS1 is found within a crater in a small catena to the south of the main Baltay catena. Unlike the previous ones, BS2 is an elongated feature, extending over ~2 km, that starts from the rim of a crater. It is oriented to the northeast and does not appear to be inside any crater nor is there a structure that accompanies it along its length.

Catenae originate from secondary impacts of fragments produced from material ejected after an initial collision. In the Yalode region, these chains extend radially from Urvara to the east and are interpreted (Crown et al., 2018) as a chain of secondary impact craters that dissect Yalode floor deposits. According to this interpretation and given the location of BS1 and the morphology of BS2, the bright spots likely are material exposed by a secondary impact.

In addition to exhibiting a red spectrum in the visible range, BS1-3 also have distinct absorption bands at 3.4 µm (Fig. 11). Although, this suggests the identification of organic material, the presence of another prominent absorption feature at 3.95-4.00 µm also suggests the existence of carbonates. It is important to note that the presence of carbonates does not imply the absence of organics. In fact, in certain regions of the Ernutet crater, there exists a direct correlation between the occurrence of carbonates and organics (De Sanctis et al., 2017).

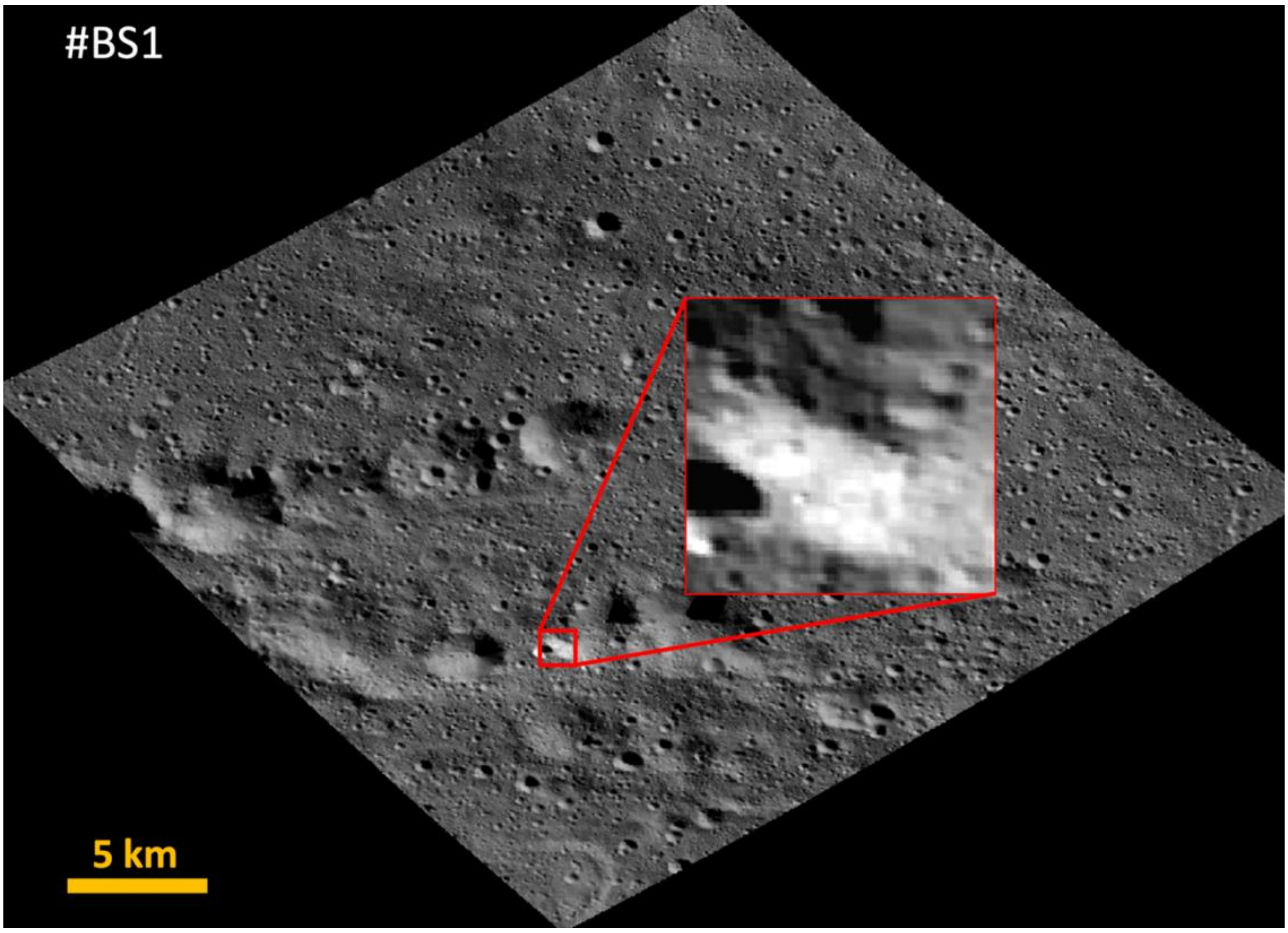
#BS1
5 km

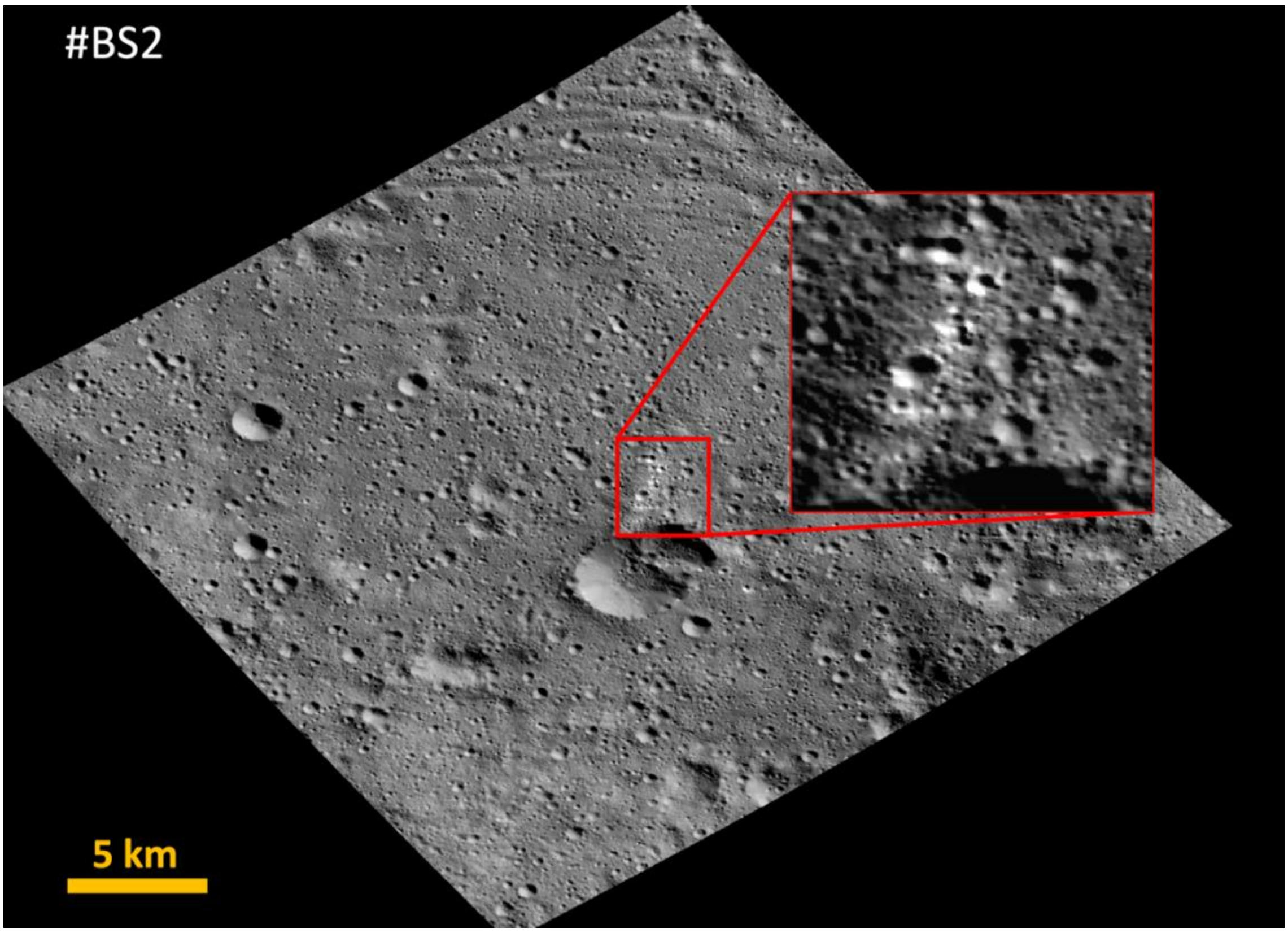
#BS2
5 km

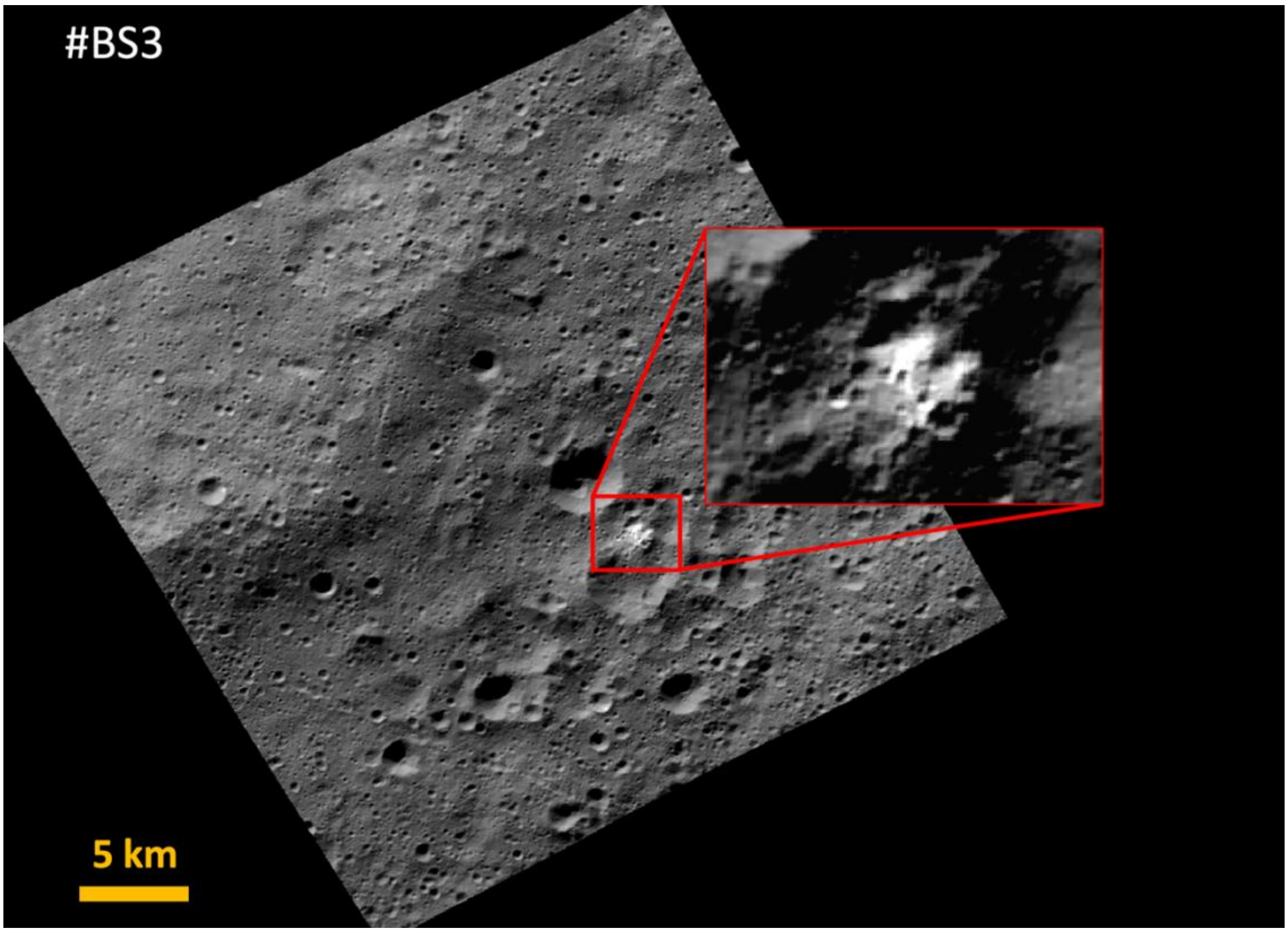

**Figure 10. Bright spots BS1, BS2, and BS3 using FC2 clear filter images, with a pixel scale of ~36 m/pixel. The bright spots extend several meters. The box on the right is a zoom of the area of interest, where the contrast has been modified for better visualization.**

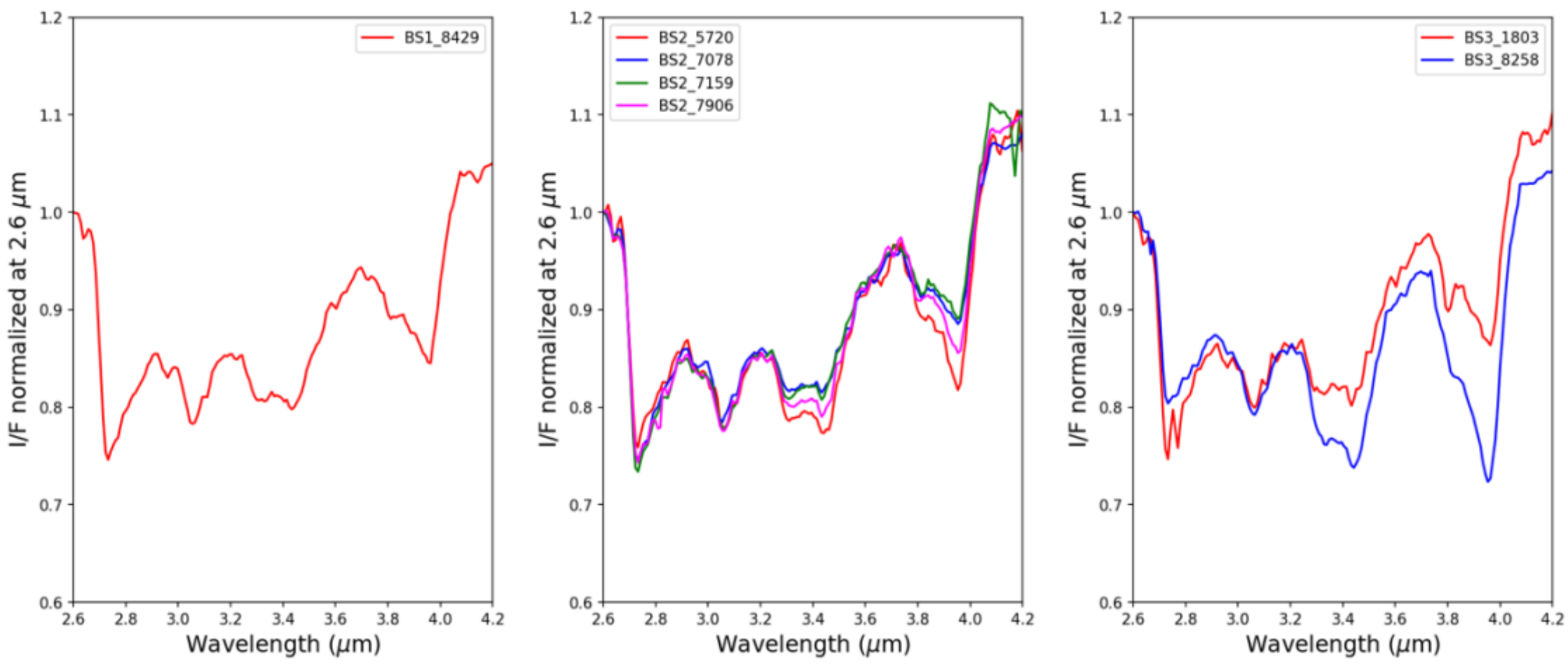

**Figure 11. a) VIR spectra for bright spots BS1, BS2 and BS3, from left to right respectively. In all cases an absorption band centered at 3.4 and 3.95 μm is present.**

Therefore, we decided to perform an analysis of the carbonates present on the surface of Ceres at a global scale to investigate the role of those identified in the region between the Yalode and Urvara basins. The objective of this approach is to compare the mineralogical features that have been identified by VIR in our BS1-3 candidates with the confirmed area containing organics, the Ernutet crater, and other areas with a high abundance of carbonates.

The data presented in Fig. 12 correspond to the absorption bands identified on a global scale in a VIR mosaic (personal communication with F. P. Carrozzo). The map displays the depth of the absorption band centered at 3.95 µm (R), the depth of the absorption band centered at 2.7 µm (G), and the center of the 3.95 µm band (B).

And indeed, this global-scale qualitative analysis allows us to group our candidate regions into bluish spots (representing most carbonates), a greenish group, which includes Ernutet, the Urvara basin (where strong evidence of organic presence has been indicated, Nathues et al. (2022), and the region under our investigation. The fact that the mineralogy exposed in BS1-3 is similar to that found in Ernutet, and differs from regions without organics, suggests that perhaps the same processes that took place in Ernutet to lead to the presence of organics may have also occurred in our region of interest.

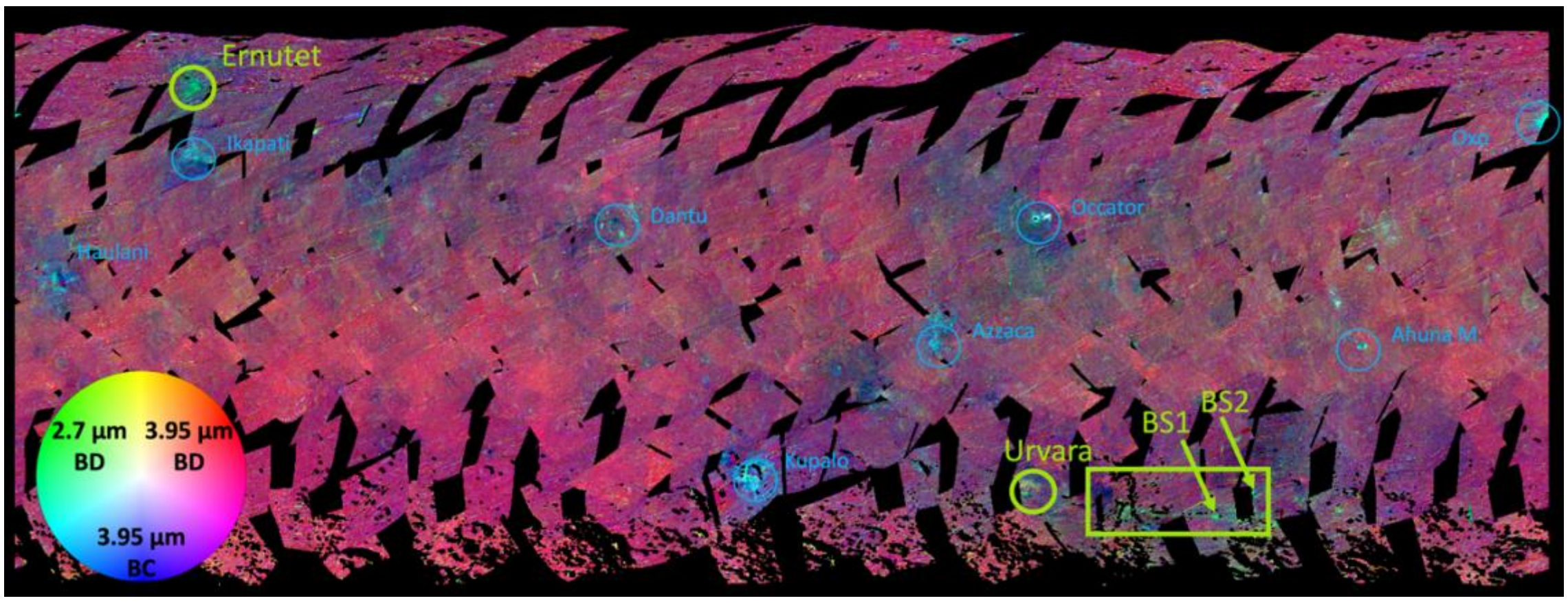

**Figure 12. Global RGB map for qualitative analysis of the role played by the carbonates identified between the Urvara and Yalode basins compared to the rest of Ceres' carbonates. We observe that this region bears a closer resemblance to the Ernutet crater and Urvara (where organics are present) than to the other carbonates-rich regions. In this figure, R corresponds to the band depth at 3.95 µm, G the band depth at 2.7 µm, and B the center of the 3.95 µm band.**

For a more quantitative analysis, we used numerical values and plotted the center of the 3.95-micron band against its depth for carbonate spots (Fig. 13, modified from Carrozzo et al., 2018). Here, Occator appears as an outlier representing carbonates with higher intensity, then a large group encompassing most of the identified carbonates on the surface, and finally a smaller separated group including the Ernutet crater, the Urvara basin, and two of our bright spots, BS1 and BS2 (BS3 is not covered by this dataset). This result indicates that the surface mineralogy observed in the Ernutet region is analogous to what we are observing in the Yalode region, particularly within the bright spots. This fact alone is not unequivocal proof of organic identification, but it is an indication of consistency.

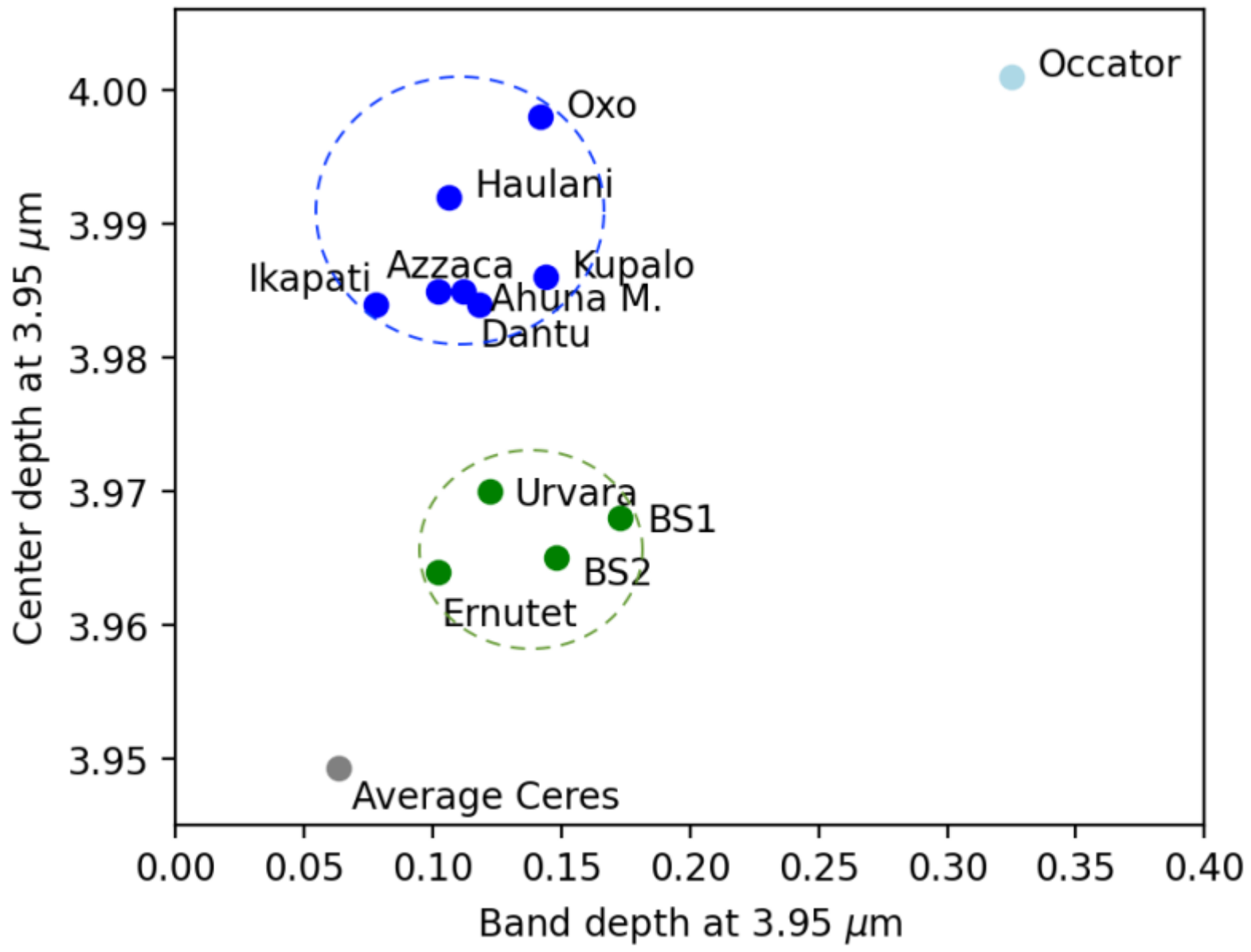

**Figure 13. Center versus depth of the 3.95-micron band (modified from Carrozzo et al., 2018). The bright spots BS1 and BS2 form a distinct group along with the carbonates from the Ernutet crater and the Urvara basin. The gray point represents the average Ceres' surface. BS3 is not in this dataset.**

## 3.5 Extrapolated dataset

The available largest pixel scale VIR data from the Yalode region do not have a uniform spatial distribution, varying from 1.103 to 0.095 km per pixel, and present large viewing geometry differences. In contrast, FC2 data present a higher uniform pixel scale over a broader area. Within this region, FC2 provides a uniform 136 meters per pixel. Consequently, it is advantageous to create spectral VIR data at the spatial scale of FC2 through an extrapolated dataset. As discussed in Section 2.2, we are assuming that the spectral behavior identified in the visible range can be extrapolated to the infrared based on the spatial correlation of VIR with a distinctive, red-sloped continuum measured by FC2 in the visible. In fact, our global survey observed that the SMA (FC2) re-detected organics in the visible range in regions where they had already been identified in the infrared by other authors (VIR). This result further validates our methodology.

To do this, first, we choose a total of 10 areas within our scene that represent the spectral diversity (Cheek and Sunshine 2020): areas with a predominance of the endmember O, zones where the average Ceres spectrum and the ejecta from small craters dominates (the endmember B), and intermediate regions with different proportions of components and shade. For each one, we calculate the VIR spectrum and the abundance of each endmember according to the previous SMA solution. Information on each of these regions, such as VIR spectrum, spatial coordinates, number of pixels occupied by FC2 and VIR data, and SMA fraction abundance, can be found in Appendix Figs. D and E.

Then, we subtract the shade from VIR spectra from the value obtained in the SMA, retrieving the shadeless spectra (Fig. 14a) to reduce the effects of topography and lightening differences between the FC2 and VIR datasets (Cheek and Sunshine 2020). Finally, we perform a least-squares fit for each wavelength using the 10 locations by assuming that the shadeless spectra are the sum of the endmember O, plus the endmember B, plus an error. The solution to this least-squares fit provides us with the spectral shape of the endmembers as seen by VIR, which are referred to as hyperspectral endmembers.

We found a hyperspectral endmember for carbonates/organics (Fig. 14b, red line) with a clear absorption centered at 3.4 µm and another at 3.95 µm, and the background Ceres spectrum (Fig. 14b, blue line) without any organics or carbonates features. In both spectra the features due to the presence of OH-bearing minerals (absorption at 2.72-2.73 µm) and ammoniated phyllosilicates (absorption at 3.1 µm) are present.

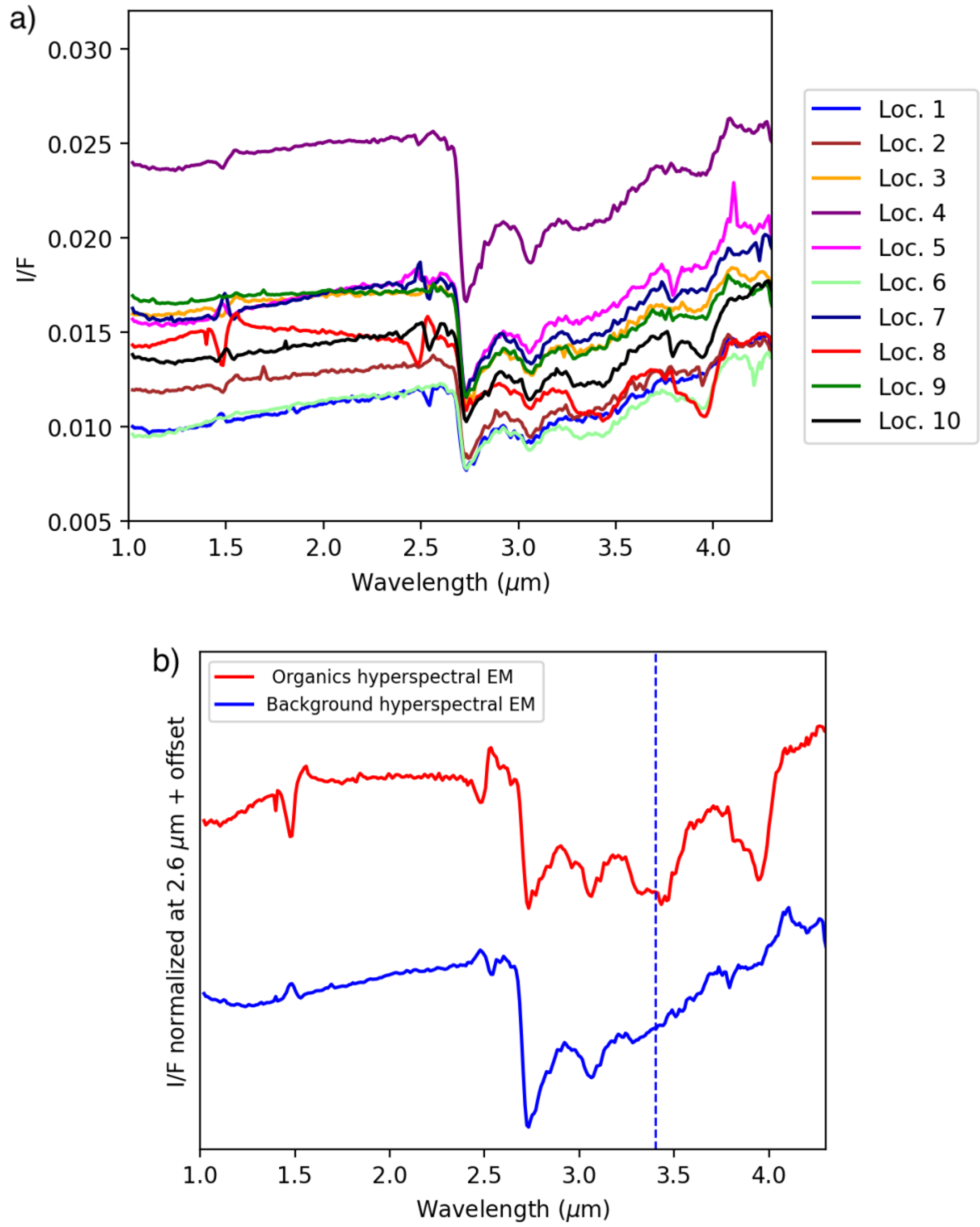

**Figure 14. a)** ***Shadeless*** **spectra from 10 locations across the scene (Yalode-Urvara region, see Appendix Fig. D for details on the locations) with different endmembers abundance. b) Hyperspectral endmembers obtained after applying Eq. (2). The red line corresponds to the carbonates/organics compound and the blue line represents the Ceres background. The blue dashed line indicates the location of the 3.4 µm absorption feature.**

The extrapolated dataset (Fig. 15) is built by summing in each FC2 pixel the endmember abundances multiplied by the hyperspectral endmember reflectance.

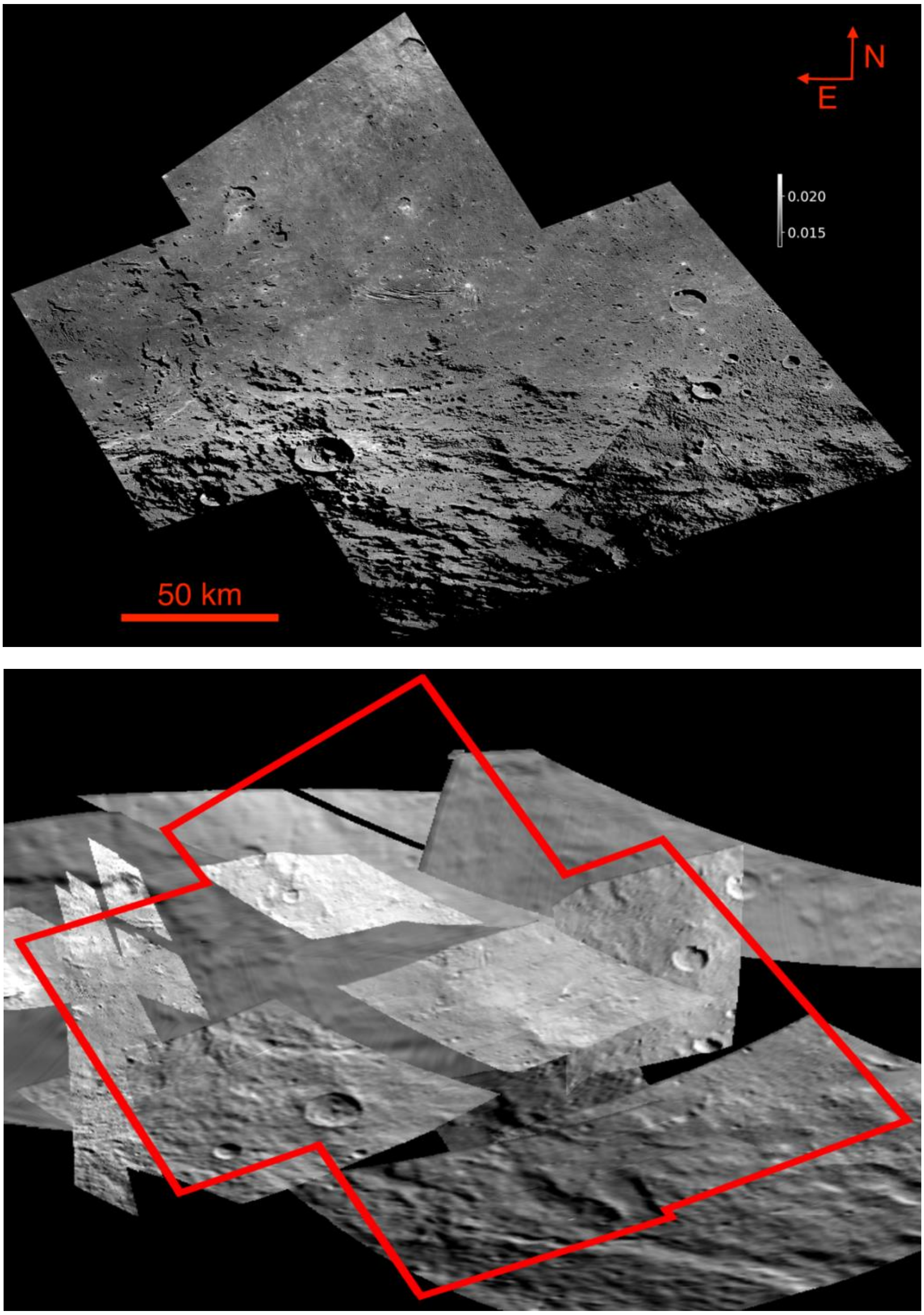

**Figure 15. Top panel: The extrapolated dataset combining FC2 (pixel scale of ~136 m/pixel) with VIR in the Yalode region. Bottom panel: Mosaic using VIR-IR data of the same region as the top panel (with the area of the extrapolated dataset outlined in red), exemplifying how the sub-maps present neither a homogeneous pixel scale nor viewing geometry. In both panels, we present the I/F at 2 μm because there are no distinct spectral features in this range.**

To characterize the presence of carbonates/organics and their spatial distribution in the scene, we calculated the depth of the band at 3.4 μm in the extrapolated dataset. First, we remove the continuum between 3.25 and 3.6 μm by fitting a straight line and dividing the reflectance. Then, the band depth is calculated by finding the minimum reflectance of a second-order polynomial fit. The result is shown in Fig. 16. In this 3.4 μm band map the bright spots identified with the SMA are spatially resolved. They have depth peaks reaching absorptions of 24, 28, and 34% for BS1, BS2, and BS3, respectively. Moreover, we find new groups of this carbonates/organics compound towards the western region of the mosaic. Additionally, we confirm the presence of a large horizontal region extending across the scene whose absorption depth is around ~4-6%. Notably, this exposure is not continuous and there are gaps with band depths that are very low.

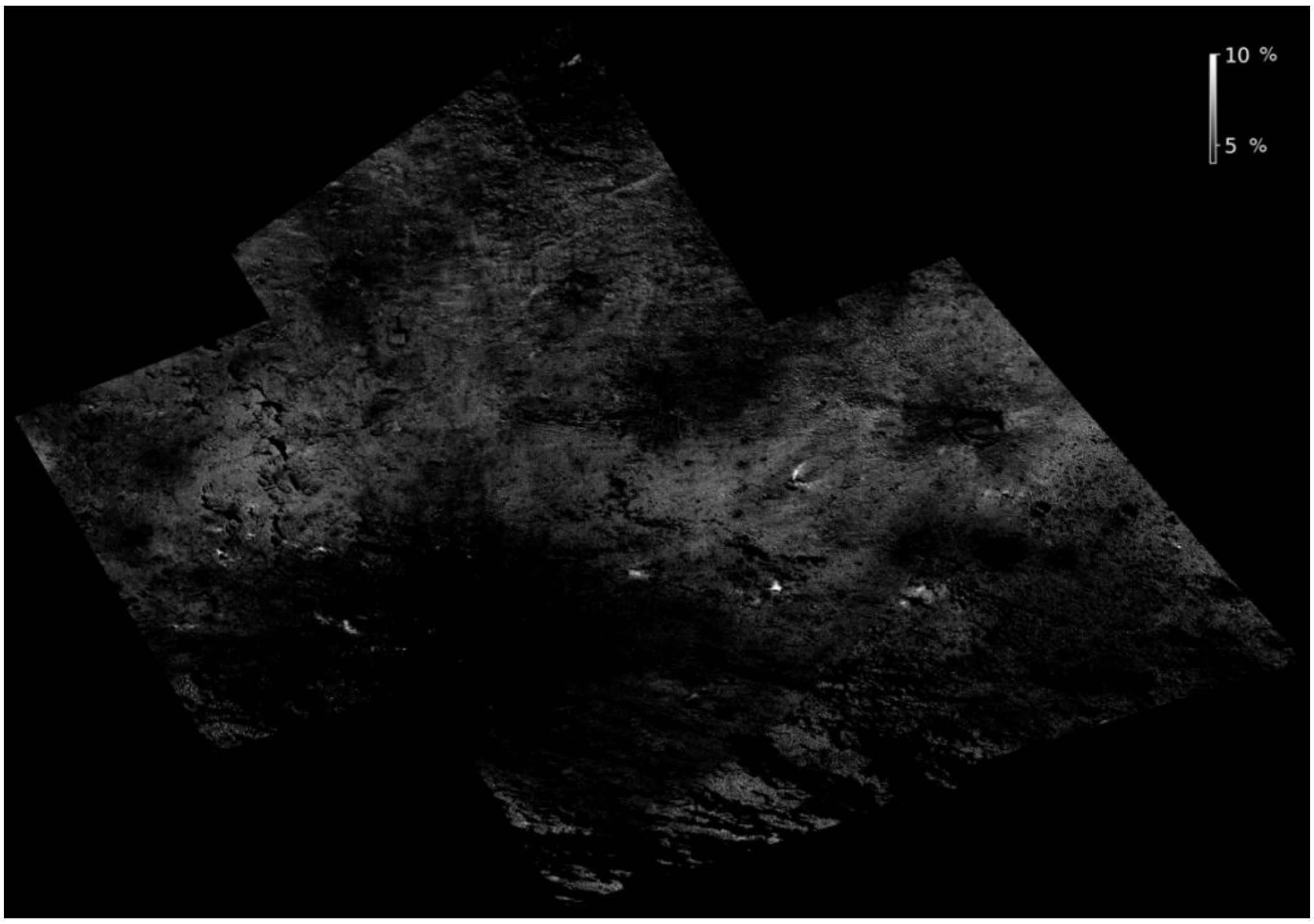

**Figure 16. 3.4 µm absorption band map computed through the extrapolated dataset. Pixel scale ~136m/pixel.**

# 4. Discussion

Our SMA analysis allows us to identify a total of 11 new candidates for organic-rich regions distributed across the surface of Ceres that exhibit red visible slopes in the FC2 images, similar to the organics found in the Ernutet crater. Other authors, using different approaches, have pointed out some of our locations. For example, Raponi et al. (2021) found hints of the presence of organic materials in the Occator crater (candidate number 10 in this work, Fig. 7i). This is a particularly interesting region, as Occator bright material is thought to come from a subsurface brine, so its confirmation would imply an endogenous nature of the organics. Nathues et al. (2022) also found that the Urvara basin could harbor organic material (candidate number 11). All the identified candidates are promising as they show red-sloped regions that are mainly concentrated on crater walls or interiors, resembling the spatial distribution of organics identified in Ernutet.

The available VIR data for 10 out of the 11 candidates lack the 3.4 µm characteristic absorption band that would confirm the presence of organics. If organics are present, interactions with solar wind and irradiation may have reduced the 3.4 absorption band. Recent studies (Daly et al., 2024) suggest that shock effects do not significantly alter the reflectance spectra of aliphatic organics in the 3.4 µm region at Ceres-like impact speeds, ruling out any type of hypothesis related to destruction of the absorption feature by subsequent impacts. Another possibility is that their abundance is low enough that we lack the necessary spatial resolution in VIR to identify them accurately. We propose these regions to be studied in future missions such as those that may one day explore the habitability of Ceres (Castillo-Rogez et al., 2022).

Nevertheless, there is one candidate where we find a clear absorption band centered at 3.4 µm, the number 12, located between the Urvara and Yalode basins. In this case, the strong presence of carbonates, indicated by the appearance of an absorption band centered at 3.95 µm, complicates the confirmation. The joint presence of carbonates and organics in Ernutet, clearly correlated in some areas, implies that there must be some mechanism favoring the presence of both components. However, in Ernutet, the presence of organics is high enough to be unambiguously identified with the VIR spectrometer data.

When comparing the absorptions of the three bright spots, BS1-3, with those of the organics identified in Ernutet, we observe that the Ernutet band is narrower (Fig. 17a). However, when we focus on the ~3.4 µm range and compare it to antigorite and dolomite – the most probable carbonate compounds for this area according to Carrozzo et al. (2018) – we see that the minor features observed in BS1 are compatible with a mixture of both the distinctive w-shaped features seen in the Ernutet organics and the dolomite absorption at 4.55 µm (see Fig. 17b). This indicates that the spectra of BS1-3 spots are compatible with the presence of the same aliphatic organics identified in Ernutet. Unfortunately, the precise nature of Ernutet organics is still unknown (De Sanctis et al., 2019), and spectral features outside the 3.4-micron region are lacking that might help disentangle them from carbonates.

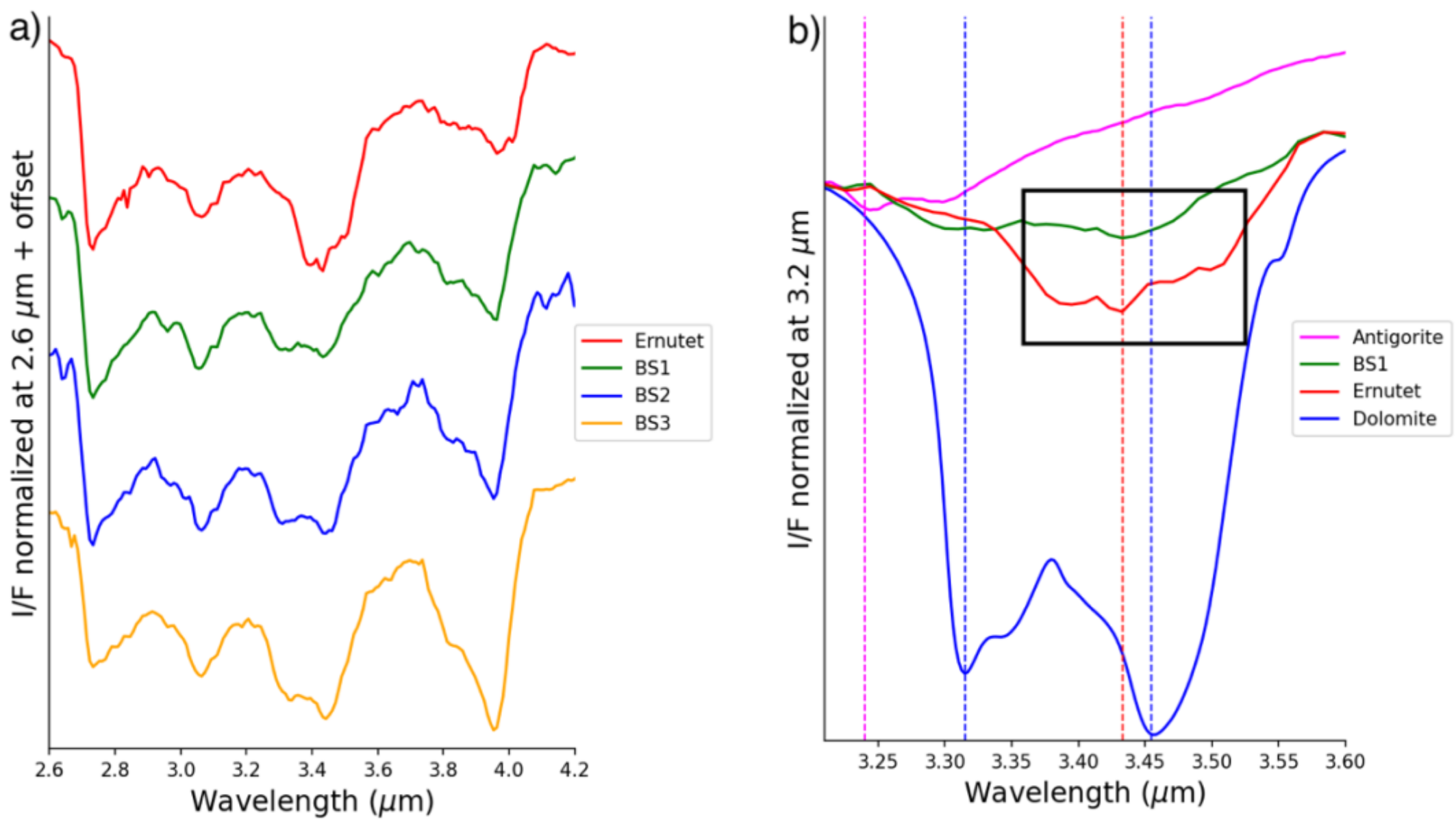

**Figure 17. a) Comparison of the spectra of the tree bright spots identified between the Urvara and Yalode basins with the organic matter identified in the Ernutet crater. Spectra are normalized at 2.6 µm and presented with an offset for clarity. b) Same as (a), but zoomed in the 3.2 – 3.6 µm range, and including the antigorite and dolomite for comparison. The dashed vertical lines indicate the most prominent absorptions. The bright spots spectra are compatible with a mixture of the organic discovered in Ernutet plus dolomite.**

Moreover, when placing this carbonate-rich region in the context of the rest of Ceres' carbonates, we find that this area closely resembles the carbonates observed in Ernutet and Urvara (Fig. 12). The differences shown in this plot are based on the center of the band around 3.95 µm, and the presence of the 2.7 µm phyllosilicates absorption centered at 2.7 µm. The role of phyllosilicates and their intimate mixture conditions are relevant as they may protect organics from the effects of space weathering by efficiently absorbing organic molecules (Poch et al., 2015; Dos Santos et al., 2016). The fact that these identified bright spots plot in a similar region as spectra from Ernutet and Urvara of Fig. 13 indicates that they might share similar mineralogical characteristics. Since mineral assemblages are indicative of formation and alteration processes, they may have undergone similar geological processes to what created the organic-bearing mineral assemblage at Ernutet, hence suggesting a common petrogenesis. Together, these considerations strengthen the case for organic-rich regions in the Urvara-Yalode region. Moreover, the fact that the stronger candidates are found beyond middle latitudes, against weaker candidates in the equatorial region, is consistent with the space weathering effects. The organics in Ernutet are located at ~51° latitude, while in Yalode, they are at ~-45°. This results in Yalode receiving only ~25% more UV irradiation than Ernutet, while the equatorial region receives up to 50% more.

The analysis of our extrapolated dataset for Yalode-Urvara (Fig. 16), allows us to focus on the spatial distribution of this material at a higher resolution. This effort revealed a wide region with 3.4 µm absorptions between the two basins Urvara and Yalode. Moreover, we identified new bright spots in the western region, in addition to the three previously identified ones, which are radially distributed with respect to the southern crater.

When analyzing the global distribution of the 3.4 µm band in the extrapolated dataset and compare with the geological units defined by Crown et al. 2018 (Fig. 18), we find that the 3.4 µm gaps from the extrapolated dataset correspond to the crater material (yellow unit), which contain the ejecta from the small craters found on the Urvara and Yalode basins. Furthermore, the region with 3.4-micron absorptions is spatially correlated to the Yalode/Urvara smooth material unit (compare Fig. 16 and 18).

According to Crown et al. (2018), the Urvara/Yalode smooth material exhibits morphological variations across its extensive expanse but is defined as a single geological unit for two reasons: the dominance of smooth surfaces, and the absence of boundaries or geological features that would support subdivision. This unit is considered to result from the coalescence or merging of deposits related to the Urvara and Yalode basins, which, because of the presence of water ice, led to the formation of this homogeneous unit. Crown et al. (2018) suspect that smooth material may represent a specific type of basin ejecta deposit, indicating the excavation of a distinct substrate and the separation of less rocky ejecta materials from the primary ejecta. The observed morphological variability in this unit is a consequence of post-formation activity, with the presence of secondary crater chains playing a role. Another contributing factor to the observed morphological variability is the variation in the underlying terrain, so areas with irregular and undulating topography are a result of the rugged subsurface material.

Based on crater size-frequency distribution measurements, Crown et al. (2018) argue that these are geological entities formed hundreds of millions of years ago; 580 ± 40 Ma for Yalode and 550 ± 40 Ma for Urvara, while the geological unit Urvara/Yalode smooth material is dated at 420 ±70 Ma. More recently, Mest et al. (2021) dated the Yalode basin to be even older (~ 1 Ga), and the Urvara basin to be significantly younger (~230 Ma). Given these ages, the organic material, if present, would have been exposed to the effects of space weathering for at least hundreds of millions of years, potentially leading to its degradation.

Yalode and Urvara were formed by the second and third largest impacts suffered by Ceres (Crown et al., 2018). Therefore, these deposits of carbonates and possible organics would come from a deeper region than the ejecta that surrounds the young and smaller craters in which this spectral feature is absent.

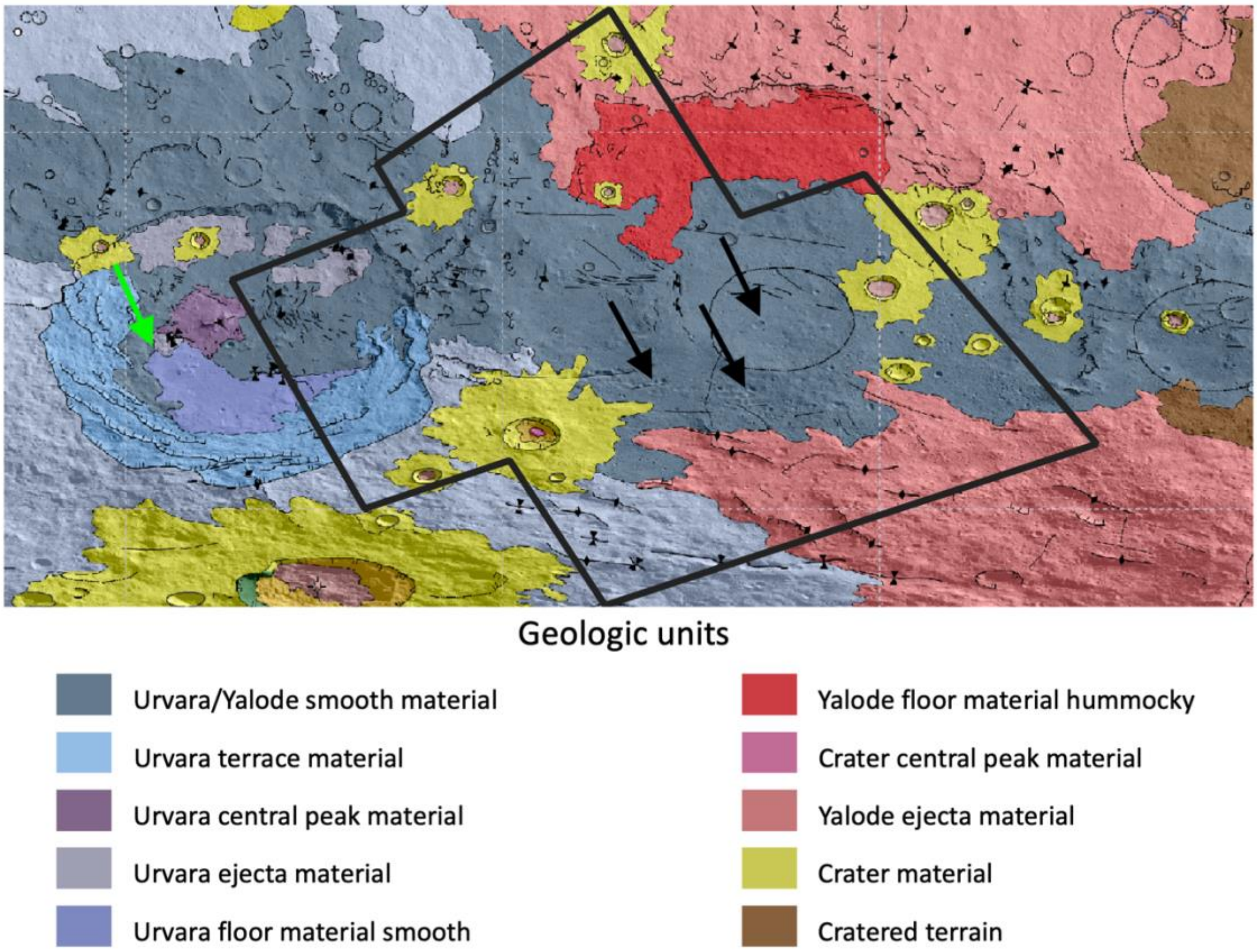

**Figure 18. Geologic units from the Yalode and Urvara quadrangles (modified from Crown et al., 2018). In black lines, we delimit the region covered by the FC2 color filters used in this work. The black arrow points toward the identified SB1, BS2, and BS3 organics candidates, all of which are located in the Urvara/Yalode smooth material. The green arrow on the left indicates the area in which the presence of organics has been proposed by Nathues et al. (2022), inside the Urvara basin.**

The fact that we encounter two overlapping craters, one old (Yalode) and one new (Urvara), where the distribution of organics is present in both, is similar to the spatial context we identified in the organic deposits in Ernutet (Fig. 5). Another region where organics have been reported is the Inamahari crater, as mentioned by De Sanctis et al. 2017. This region is an intriguing geological feature characterized by a system of two craters of similar size that are superimposed on each other. The first crater, likely formed by an earlier impact event, exhibits typical features such as a well-defined rim and a central peak. The subsequent impact that created the secondary crater partially overlapped the primary crater, resulting in a complex topography where features from both craters are visible. This suggests a mechanism in which the presence of a significant-sized ancient crater combined with a younger impact plays a crucial role in the excavation mechanisms proposed by De Sanctis et al. (2019) and supports a deeper endogenous origin of these materials. However, Ceres' surface is filled with ancient craters overlapping younger ones where no organics have been observed. Therefore, this could be a necessary but not sufficient condition.

Moreover, the new map of Ernutet (Fig. 4) shows granularity in organic materials. These materials are found indiscriminately in the outer region of Ernutet, on its walls, inside, as well as in the younger interior craters. There is no apparent spatial relationship between large impact features and this overall distribution, and although this does not rule out the exogenous hypothesis (Pieters et al. 2018), it suggests that it can be interpreted as a sign that it is endogenous material that has surfaced. In addition, the diffuse organic spot outside of Ernutet (Fig. 5), which seems to be correlated with an ancient crater, includes a large number of small craters. These small impacts may have contributed in some way to exposing the subsurface material. Confirmed organics in Ernutet and candidates in the Urvara-Yalode region are dispersed over tens of kilometers. However, unlike the 36 m pixel scale of Ernutet, the best resolution for the Yalode-Urvara region is ~136 m/pixel, therefore it is not possible to make a proper comparison of the granularity at Ernutet versus these other areas.

The most interesting distribution of bright spots is found in BS2 (Fig. 19). The feature is located in the vicinity of a crater but extends outward. It exhibits a distribution resembling a horseshoe, with its tips extending predominantly to the northwest. When comparing it with the image of the same region obtained through the clear filter FC2, we find no obvious indications of an ancient crater or a relationship with any other geological structure throughout its extent.

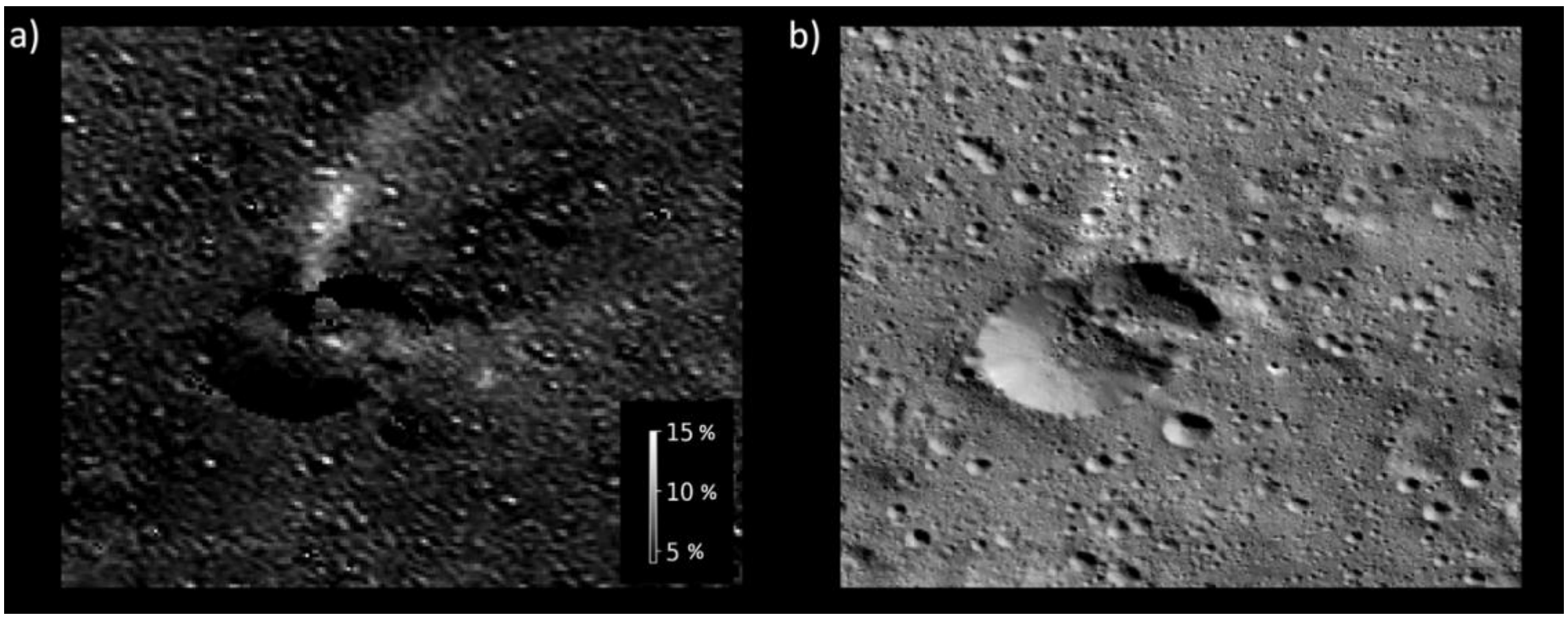

**Figure 19. a) Same as Fig. 16 but zoomed in on BS2, showing the distribution of the 3.4 absorption band of the possible organics. This material has a horseshoe-shaped distribution. b) Same area as seen by the FC2 clear filter for contextualization. No ancient crater or geological feature is aligned with this distribution.**

This distribution preliminarily led us to consider that the material in this location had been delivered by an impact. However, features like the Baltay or Pongal catenas (Fig. 8a) account for significant geological activity during and after the impacts that originated it. Baltay catena is believed to be an impact crater chain that originated after the Urvara or Yalode impacts. The mechanism that formed Pongal Catena is not understood yet, but it is believed to be a consequence of the two impacts as well. In this area, there are also channels and grooves, which suggest the collapse of the surface materials (Crown et al. 2018). Thus, taking into account that BS1 is located in a catena, the elongated morphology of BS2, the radial distribution of the bright spots, and the relevant geological activity that has been developed in this area (impacts, stresses, and deformations), the identified bright spots are compatible with fresher Yalode/Urvara smooth material exposed by subsequent impacts. Thus, this suggests that it would have an endogenous origin and is consistent with the hypothesis of De Sanctis et al. (2024), whose recent laboratory experiments lead them to conclude that organics must be present in substantial quantities within the subsurface.

Considering all the evidence presented, our findings suggest that an endogenous origin of the organics on Ceres is more plausible. This material may have surfaced through mechanisms proposed by De Sanctis et al. (2019) based on Ceres' internal evolution (Castillo-Rogez et al., 2018; McCord & Castillo-Rogez, 2018). The transfer of material within Ceres is expected to be unidirectional toward the surface, facilitated by the likely pressure of the residual brine layer or reservoir (Neveu & Desch, 2015). However, further evidence is needed to confirm this hypothesis, and an exogenous origin cannot yet be completely ruled out.

The widespread presence of ammoniated phyllosilicates and carbon on the surface suggests that Ceres likely originated at larger heliocentric distances and later than CI chondrites, in a region of the solar system where initial thermodynamic conditions allowed for the efficient accretion of significant amounts of nitrogen and carbon (De Sanctis et al., 2015, Zolotov et al., 2020; Castillo-Rogez et al., 2022). The idea that Ceres formed in distant regions, such as the Kuiper Belt, was already suggested by McKinnon, 2018. If this hypothesis holds, then an endogenous origin of the organics implies that these compounds could have also been formed in the outer solar system regions. This notion aligns with Capaccioni et al.'s (2015) proposal, suggesting that the higher abundance of organics on the surface of the comet 67P/Churyumov-Gerasimenko comet, compared to other Jupiter Family Comets (JFCs), might be linked to a formation scenario occurring at low temperatures in distant regions like the Kuiper Belt.

# 5. Conclusions

We characterized the organics observed in the Ernutet crater with FC2, to globally evaluate the potential presence of aliphatic organics in other regions using Spectral Mixing Analysis (Adams & Gillespie 2006). By applying this technique first to the Ernutet data with the highest spatial resolution, we generated a mosaic that reveals a granular spatial distribution of organic-rich material. These organics are present indiscriminately in the outer region of Ernutet, on its walls, interiors, and even within the younger inner craters. Moreover, we found that the highest concentrations of aliphatic organics show a circular pattern and seem to be correlated with an ancient crater in which Ernutet has been superimposed. The association of high concentrations of organics with successive small impacts suggests that the organic material may originate from depth.

Globally, SMA also allows us to identify 11 new candidate regions with potential organic material with a noticeable concentration in the equatorial region. There is no association between these candidates and high concentrations of carbonates or phyllosilicates. Out of those 11 candidates, 10 lack organic absorptions in VIR data and cannot be confirmed to be organic-rich, perhaps because the characteristic 3.4 µm absorptions are suppressed by space weathering and/or are below the detectability limit in the lower spatial resolution VIR data. However, in one of the candidates, VIR spectra do unambiguously detect the 3.4 µm absorption band—characteristic of carbonates and/or organics— in some bright spots at the quadrangle Yalode, situated between the Urvara and Yalode basins.

Focusing on the Yalode-Urvara region, we detect a high concentration of carbonates. The carbonates found in that location are similar to those observed in Ernutet and Urvara, suggesting that the mineral assemblage present at Ernutet is also present at the bright spots, potentially indicating a shared origin or petrogenesis for the materials in both areas. This contrasts with other carbonates in regions where no evidence of organic material has been detected. Furthermore, inspection of VIR data between 3.2-3.6 µm shows that minor spectral features of the bright spots align with a mixture of the Ernutet organics and dolomite, the carbonate believed to be more abundant in this region. Therefore, the Yalode region is a strong candidate to contain organic matter. However, if present, it may be in too low a concentration to be unambiguously confirmed.

By creating an extrapolated dataset through SMA—combining the spatial resolution of FC2 with the spectral resolution of VIR— we studied the spatial distribution of the 3.4-µm absorption band in the Yalode and Urvara basins. The strength of the 3.4 µm band correlates well with what Crown et al. (2018) defined as Yalode/Urvara smooth material, a geological unit, that emerged following the second and third most significant impacts suffered by Ceres. The spatial distribution of this material is consistent with the idea that the brightest spots (that is, the areas with the highest abundance of organic-rich endmember) are regions of fresher material exposed because of the significant geological activity in the area. Thus, at Urvara/Yalode, as in Ernutet and Inamahari, dual systems of new and old craters are superimposed, which may indicate that an impact on an old crater facilitates the movement of these deeper organics up toward the surface.

The association of organic-rich materials with complex and multiple large-impact events, strongly suggests that the organics on Ceres are endogenous. Supporting evidence for an endogenous origin of the organics, based on numerous detections of sites with organic-rich material, would point to the presence of an organic-rich reservoir within Ceres, which bears significant astrobiological implications and may affect priorities in the coming decades of solar system exploration.

Future lander missions to Ceres hold the promise of shedding light on the intriguing discovery of organics on its surface, providing a more definitive understanding of their nature. *In situ* sample return exploration of Ceres in the next decade would be complementary to the previous sample return missions to the primitive asteroids Bennu and Ryugu –OSIRIS-REx (Lauretta et al., 2017) and Hayabusa-2 missions (Tsuda et al., 2013)–, and to the future ocean worlds in the outer solar system –Europa Clipper (Daubaret al., 2024) or Jupiter Icy Moons Explorer missions (Fletcher et al., 2023). While current observations provide suggestive hints of organics in regions other than Ernutet, uncertainties remain regarding their nature and composition. Upcoming missions equipped with advanced instrumentation and capabilities could perform high-resolution, detailed analyses of these materials and confirm whether the presence of organics occurs on the entire surface or not.

In a related context, NASA's Lucy mission (Emery et al., 2024b), set to explore the Trojan asteroids, could offer valuable insights into the distribution and characteristics of organic compounds in the solar system. Specifically designed to detect organics (Levinson et al., 2021; Reuter et al., 2023), Lucy's exploration of these ancient, pristine bodies in the Sun-Jupiter Lagrange points may provide valuable insights into the nature and origins of organic materials in the early solar system. While Trojan asteroids and Ceres are distinct objects with different histories, any unexpected findings of organic compounds on the Trojans could broaden our understanding of the processes that govern organic chemistry in diverse environments. Though the direct comparison between Trojans and Ceres is limited, Lucy's findings might still offer indirect contributions to our understanding of organics on other solar system bodies, including Ceres, and their implications for the broader search for life beyond Earth.

# Appendix

**Figure A. Fractional abundance maps obtained after applying the SMA in the FC2 mosaic over the Ernutet area.**

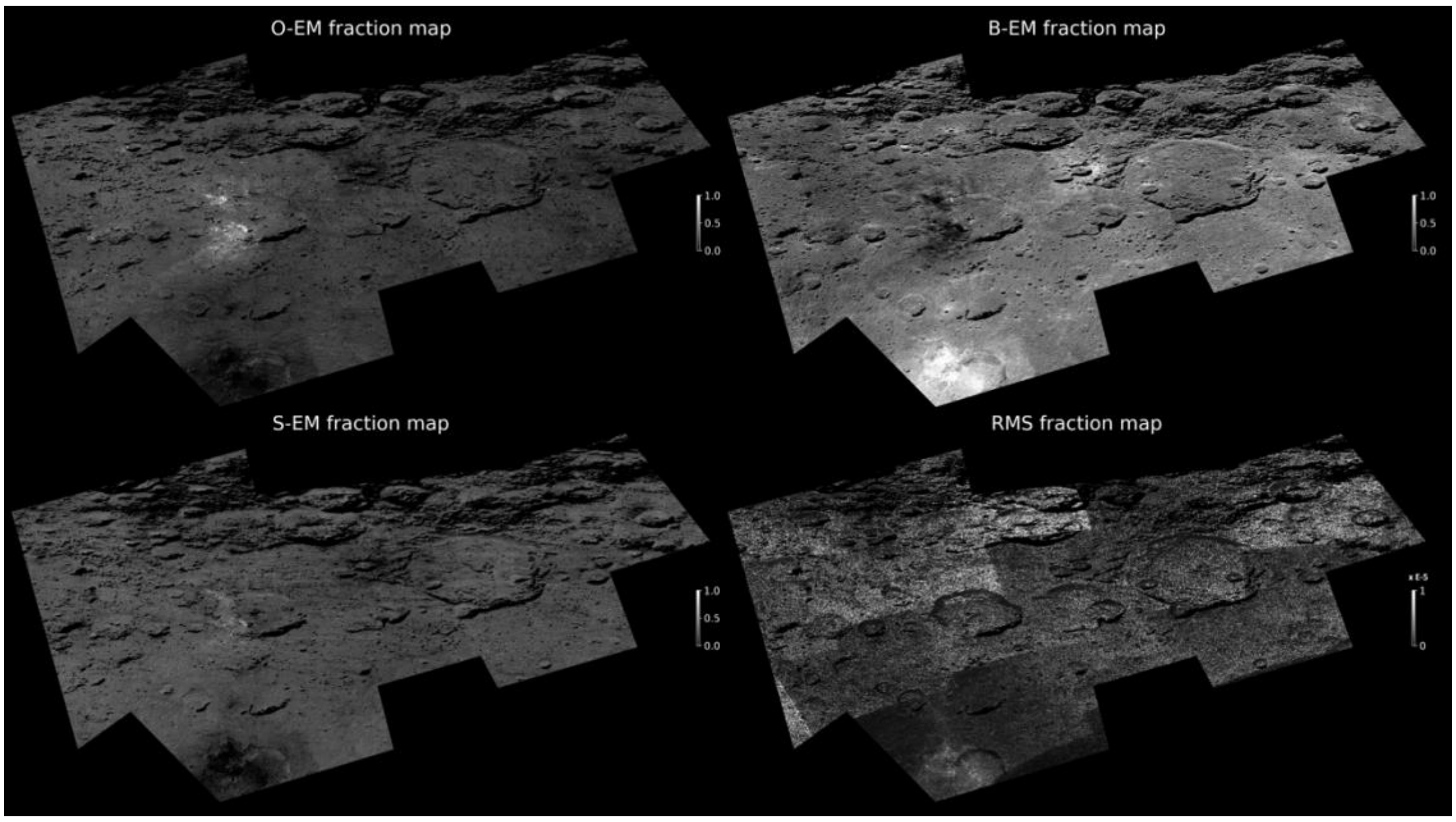

**Figure B. Fractional abundance maps obtained after applying the SMA in the FC2 mosaic over the XMO images from the Ernutet region.**

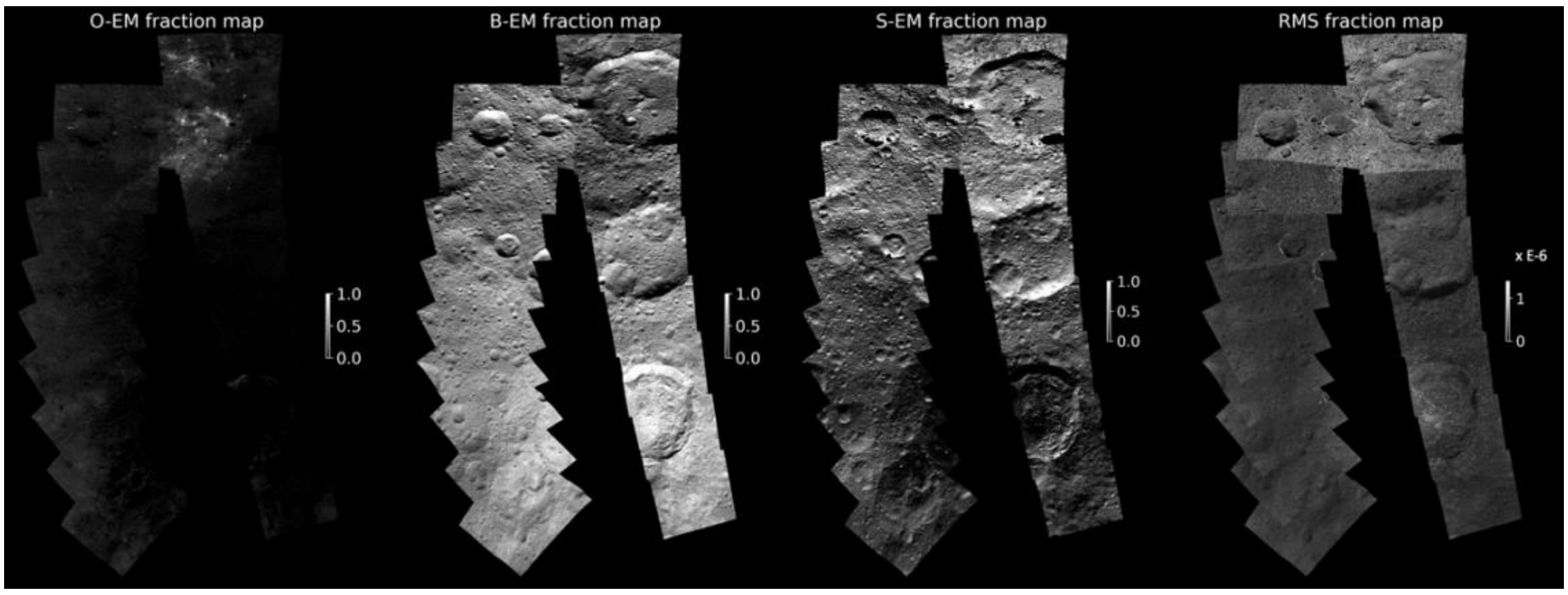

**Figure C. Fractional abundance maps obtained after applying the SMA in the FC2 mosaic over the Yalode and Urvara basins.**

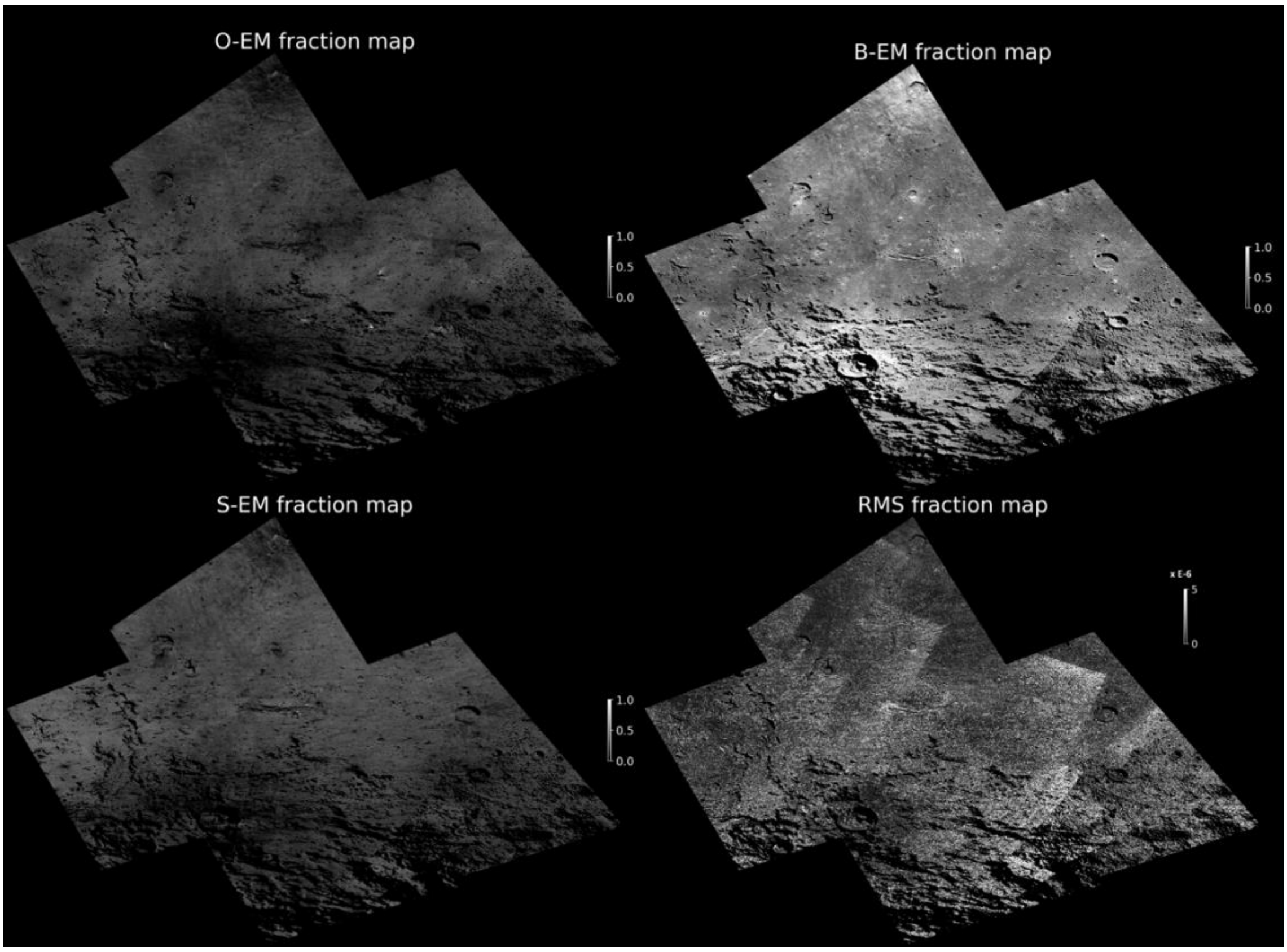

**Figure D. Color composite of three fractional abundance maps of the Yalode-Urvara region. Locations for representative spectra (presented in Fig. E) are shown.**

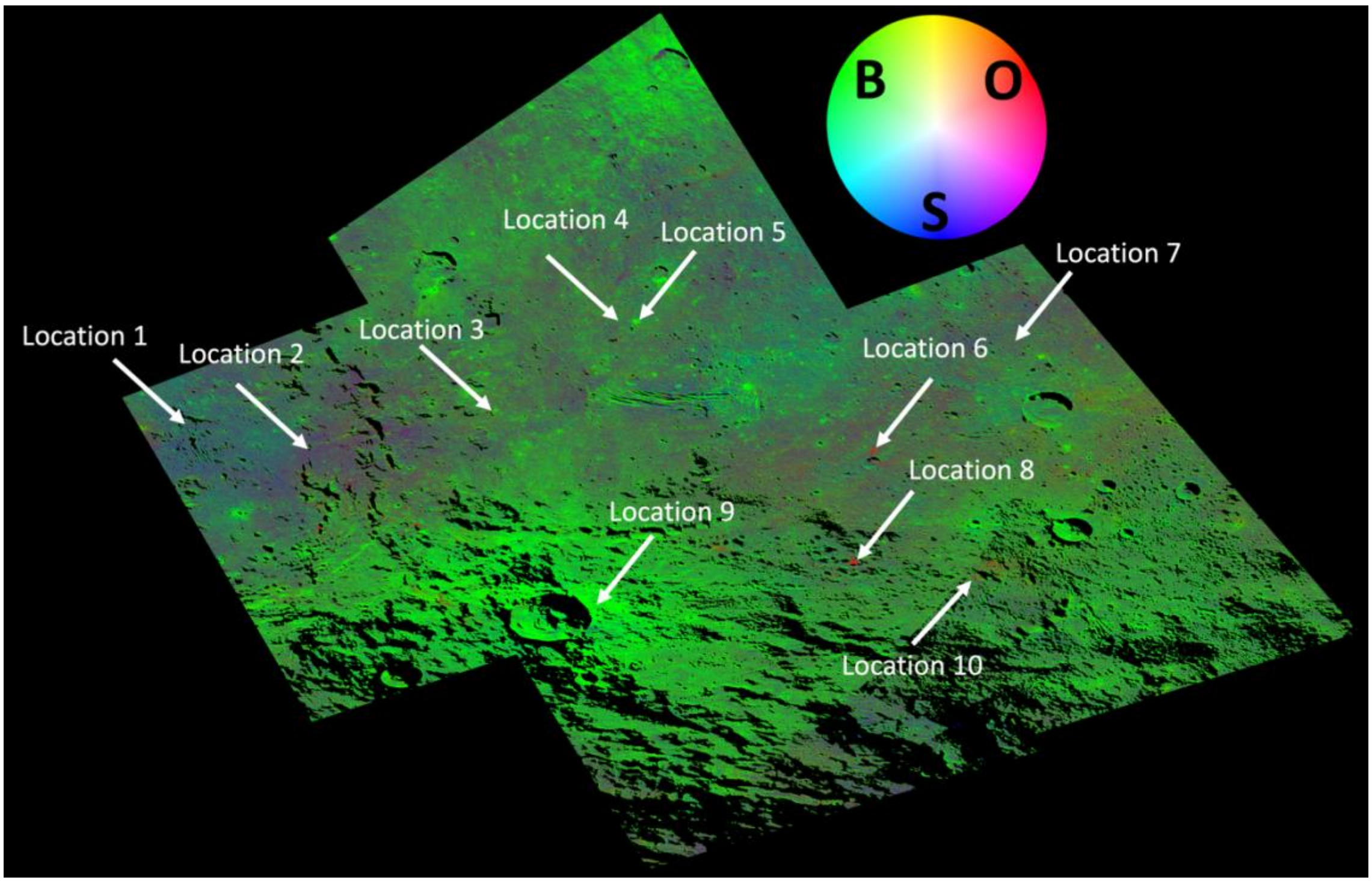

**Figure E. Measure VIR spectra for each one of the locations indicated in Fig. D. Each plot includes the number of pixels occupied by both FC2 and VIR, the fractional abundance of each endmember, and the geocentric coordinates. The vertical blue line is located at 3.4 µm.**

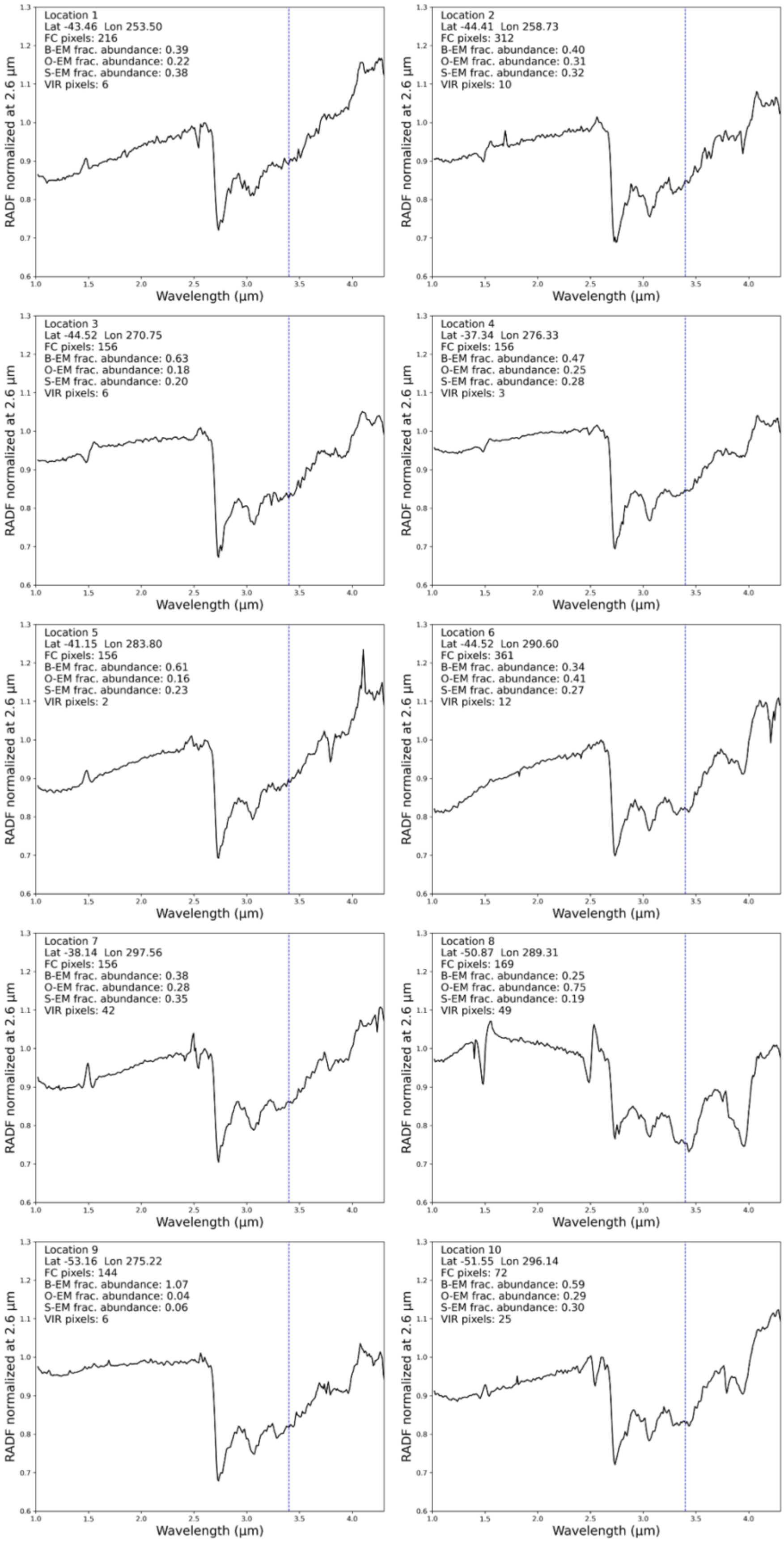